\documentclass[structabstract, english]{aa}  
\pdfoutput=1
\usepackage{amsmath}
\usepackage{graphicx}
\usepackage{txfonts}
\usepackage{natbib}
\usepackage{subfig}
\usepackage{booktabs}
\usepackage{upgreek}
\usepackage{setspace}
\usepackage{gensymb}
\usepackage{lscape}
\usepackage{lipsum, babel}
\usepackage{comment}
\usepackage{multirow}
\usepackage{changepage}
\usepackage{float}
\usepackage{appendix}
\usepackage{longtable}
\usepackage{xtab,booktabs}
\usepackage{tabularx}
\usepackage{epsfig}
\usepackage{morefloats}

\newcommand{\C}{\textsc{cloudy}}
\newcommand{\Sh}{\textsc{shape}}

\begin{document}
%

   \title{Polarimetry and Spectroscopy of the `Oxygen Flaring' DQ Herculis-like nova: V5668 Sagittarii (2015)}


   \author{E. J. Harvey
          \inst{1},
          M. P. Redman\inst{1}, M. J. Darnley \inst{2}, S. C. Williams \inst{2}$^{,}$\inst{3}, A. Berdyugin \inst{4}, V. E. Piirola \inst{4}, K. P. Fitzgerald \inst{1}, E. G. P. O' Connor \inst{1}}

\institute{Centre for Astronomy, School of Physics, National University of Ireland Galway, University Road, Galway, Ireland 
\email{e.harvey2@nuigalway.ie, matt.redman@nuigalway.ie} 
\and{Astrophysics Research Institute, Liverpool John Moores University, IC2, Liverpool Science Park, Liverpool, L3 5RF, UK}
\and{Physics Department, Lancaster University, Lancaster, LA1 4YB, UK}
\and{University of Turku, Tuorla Observatory, Vaisalantie 20, 21500 Piikkio, Finland}
}
    
   \date{\today}



  \abstract
{Classical novae are eruptions on the surface of a white dwarf in a binary system. The material ejected from the white dwarf surface 
generally forms an axisymmetric shell of gas and dust around the system. The three-dimensional structure of these shells is difficult 
to untangle when viewed on the plane of the sky. In this work a geometrical model is developed to explain new observations of the 
2015 nova V5668 Sagittarii.}
{We aim to better understand the early evolution of classical nova shells in the context of the relationship between polarisation, photometry and spectroscopy in the optical regime. To understand the ionisation structure in terms of the nova shell morphology and estimate the emission distribution directly following the light-curve's dust-dip.}
{High-cadence optical polarimetry and spectroscopy observations of a nova are presented. The ejecta is modelled in terms of morpho-kinematics and photoionisation structure. }
{Initially observational results are presented, including broadband polarimetry and spectroscopy of V5668 Sgr 
nova during eruption. Variability over these observations provides clues towards the evolving structure of 
the nova shell. The position angle of the shell is derived from polarimetry, which is attributed to scattering from small dust 
grains. Shocks in the nova outflow are suggested in the photometry and the effect of these on the 
nova shell are illustrated with various physical diagnostics. Changes in density and temperature as the super soft source phase 
of the nova began are discussed. Gas densities are found to be of the order of 10$^{9}$ cm$^{-3}$ for the nova in its auroral phase. 
The blackbody temperature of the central stellar system is estimated to be around $2.2\times10^{5}$ K at times coincident 
with the super soft source turn-on. It was found that the blend around 4640 $\rm{\AA}$ commonly called `nitrogen flaring' 
is more naturally explained as flaring of the O~{\sc ii} multiplet (V1) from 4638 - 4696 $\rm{\AA}$, i.e. `oxygen flaring'. 
}
{V5668 Sgr (2015) was a remarkable nova of the DQ Her class. Changes in absolute polarimetric and spectroscopic multi-epoch observations lead to interpretations of physical characteristics of the nova's evolving outflow. The high densities that were found early-on combined with knowledge of the system's behaviour at other wavelengths and polarimetric measurements strongly suggest that the visual `cusps' are due to radiative shocks between fast and slow ejecta that destroy and create dust seed nuclei cyclically.}

\authorrunning{Harvey et al.}
\titlerunning{}
	\maketitle

\section{Introduction}

Classical novae are a sub-type of cataclysmic variable and are characterised by light-curves and spectra whose development are followed from radio through to gamma wavelengths. 
\cite{Strope:2010aa} classified 
a variety of optical light curves and provided physical explanations for many of their features and more 
recently \cite{Darnley_prog} laid out a new classification scheme for novae based on the characteristics of the companion star. 


Novae are known to be a distinct stellar event and in their simplest terms are considered as either fast (t$_3$ $\textless$ 20 days) or slow (t$_3$ $\textgreater$ 20 days, where t$_3$ is the time taken for a nova's magnitude to decrease by 3). Fast novae occur on more massive white dwarfs than slow novae and require less accreted matter in order to ignite the thermonuclear runaway and experience higher ejection velocities, e.g. \cite{yaron}. Slower novae counterparts 
typically occur on lower mass white dwarfs, eject more previously-accreted-material during eruption and the outflow has lower ejection velocities which creates rich dust formation factories, e.g. \cite{evanswise}. These objects are well observed during eruption where optical photometry and spectroscopy are the most thoroughly practiced approaches. Although in recent times X-ray observations have become more common with the advent of $\it{Swift}$, see \cite{swiftxray}.

Novae have long been observed in terms of spectroscopy dating back to the 1891 nova T Aur \citep{Taur_disc}, and have been systematically studied since \cite{tololo}. A user guide on spectroscopy of classical novae is also available \citep{shoreuser}. A commonly adopted classification scheme for nova spectroscopy is known as the Tololo scheme, first presented by \cite{1991ApJ...376..721W}. Nova spectra are characterised by several observable stages during their eruption and the progression through these spectral stages (i.e. skipping some or showing critical features in others) gives the nova its spectral fingerprint. The spectral stages in the Tololo scheme are defined by the strength of the strongest non-Balmer line, as long as the nova is not in its coronal stage (given the designation C, defined as when [Fe~{\sc x}] 6375 $\rm{\AA}$ is stronger than [Fe~{\sc vii}] 6087 $\rm{\AA}$), whether they are permitted lines, auroral or nebular (P, A or N respectively). Depending on which species is responsible for the strongest non-Balmer transition in the optical spectrum, the formulation (h, he, he$^{+}$, o, ne, s...) is denoted by a subscript. At any time, if the O~{\sc i} 8446 $\rm{\AA}$ line is present then an `o' is included as a superscript in the notation. Developing spectral stages of novae are also often described as the pre-maximum spectrum, principal, diffuse enhanced, orion, auroral and nebular (in order of appearance) see e.g. \citet{warner,anu2012}. Changes in the appearance of nova spectra are due to temperature, expansion, clumping, optical depth effects, contribution from the companion star and orbital phase.  

To date, polarimetric observations of novae have shown intrinsically low levels of absolute polarisation and are therefore 
difficult to quantify and understand. Polarimetric observations of novae began over fifty years ago during early development of 
the technique as it applies to astronomical objects. V446 Her was the first nova to be observed and a constant linear polarisation 
was measured to within 0.13$\%$ in absolute polarisation degree \citep{grigorian}.

Observations of novae tend to demonstrate low polarisation arising from several astrophysical processes such as clumpiness, 
scattering by small dust grains, electron scattering and polarisation in resonance lines, or a combination of these processes. 

The dust formation episode in novae is identified as a deep dip in the visual light curve, corresponding to a rise in the thermal infrared, known commonly as the `dust-dip'. As the newly-formed optically-thick dust shell expands away from the central system the visual brightness increases once again, and although the recovery is often smooth it is possible to have cusp-shaped features in this part of the light-curve that are often associated with radiative shocks in the ejecta \citep{lynch_cusps,kato2009}. These shocks are expected in part to contribute to the ionisation of the nova ejecta as well as the shaping, clumping, dust formation and destruction processes. Shocks are detectable in the radio, gamma and X-ray wavelength regimes \citep{metzger}. The role that shocks have in the early evolving nova outflow has been analysed in detail by \cite{derdzinski}, who found that consequences of a shock treatment over a purely homologous photoionised expansion lead to higher densities and lower temperatures in certain parts of the ejecta.

DQ Her is an historically important nova-producing system.  Following a major observed eruption in 1934, DQ Her became the archetype for rich dust-forming slow novae. It was one of the first nova to be followed with high-cadence spectroscopic observations \cite{strattonDQher}, and this data was later used to classify nova spectra into 10/11 subclasses by \cite{dbmclaughlin_class}. \cite{walker54} showed a binary and since then it has been established that all classical nova-producing systems contain binary cores. \cite{kemp_linear_dqher} found variation in linear polarisation of the quiescent DQ Her system and \cite{kemp_circular_DQHer} presented variations in the circular polarisation of DQ Her. Both the circular and linear polarisation variation were found to correspond to twice the white dwarf period of 71 s. In a later paper, \citet{penning_circularpol} found no variation in the circular polarisation of DQ Her that corresponded to the white dwarf's orbital-spin period, although no reference was made to the work of either \cite{kemp_linear_dqher} or \cite{kemp_circular_DQHer}.

\begin{figure}[ht!]
\includegraphics[width=9cm]{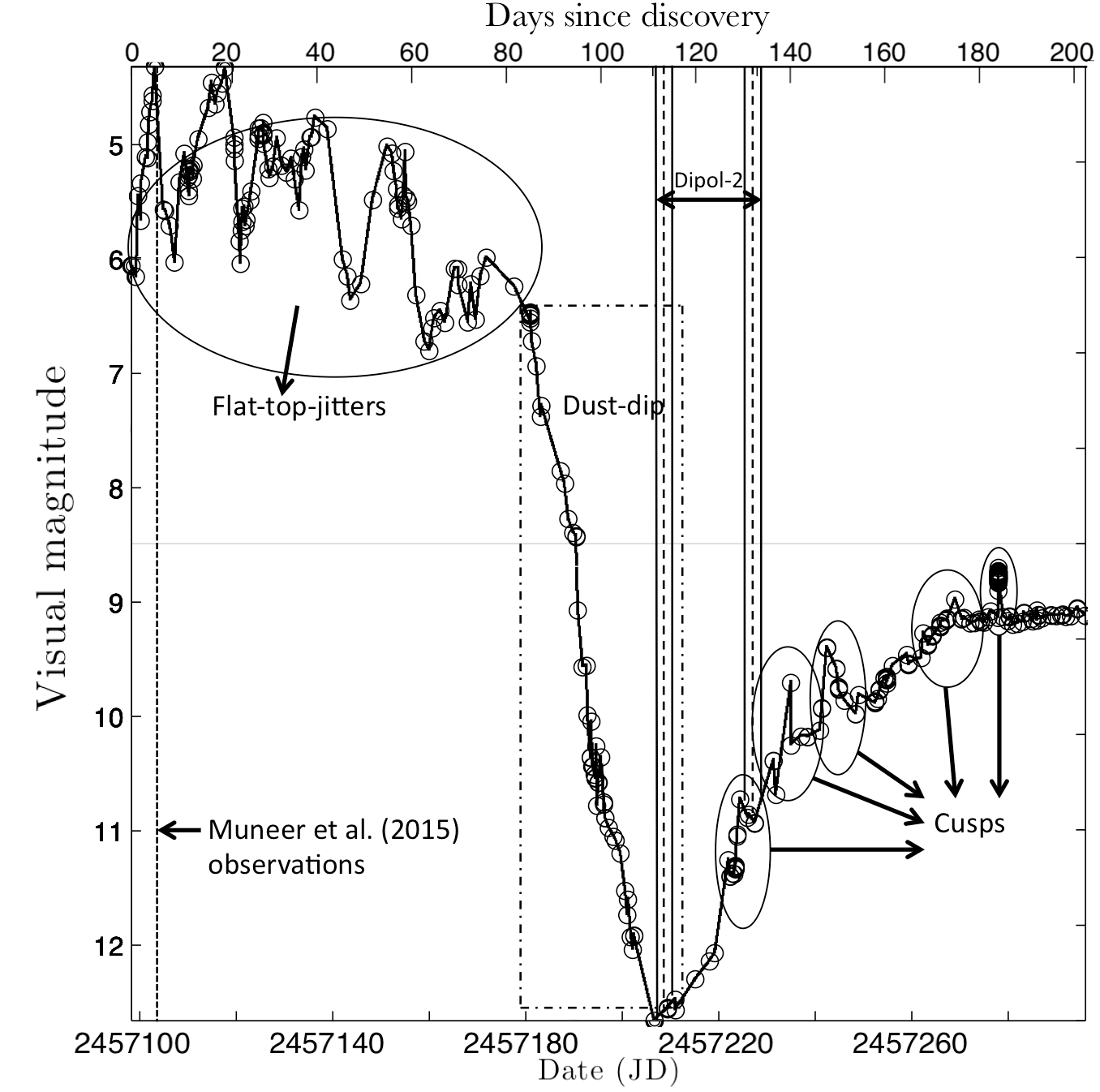}
\caption{V5668 Sgr AAVSO light curve. Marked are the three major light-curve stages observable in the figure, i.e. the flat-top-jitters, the deep dust-dip and the cusp shaped features seen on the rise out of the dust-dip. Marked are the times of polarimetric observations of the nova by both the \cite{muneer} team and those presented in this paper, i.e. the Dipol-2 measurements. The y-axis demonstrates the change in visual magnitude whereas the x-axis contains the Julian date on the bottom and days since discovery on the top.}
\label{fig:V5668Sgr_LC}
\end{figure}

The subject of the paper is V5668 Sgr (PNV J18365700-2855420 or Nova Sgr 2015b) a slow-evolving dust-forming nova and is a clear example of a DQ Her-like nova. V5668 Sgr was confirmed as an Fe~{\sc ii} nova in spectra reported by \cite{Atel7230} and \cite{Atel7265} after it was discovered at 6.0 mag on 2015 March 15.634 \citep{seach}. As a close and bright nova with a deep dust-dip, this object might be expected to produce a visible shell discernible from the ground within ten years using medium class telescopes.  \citet{banerjeeV5668sgr} calculated a distance of around 1.54 kpc to the nova system. The distance was calculated by fitting an 850 K blackbody to their dust SED on day 107.3 post-discovery to find $\rm{\theta _{bb}}$ and assuming an expansion velocity of 530 km s$^{-1}$, where $\rm{\theta _{bb}}$ is the blackbody angular diameter. It was also found in the same work that V5668 Sgr was a rich CO producer as well as one of the brightest novae (apparent magnitude) of recent times, reaching 4.1 mag at visual maximum. As V5668 Sgr is a clear example of a DQ Her-like nova light curve (see Fig. \ref{fig:V5668Sgr_LC}), it is interesting to look for similarities between the two systems. A 71 $\pm$ 2s oscillation in the X-ray flux was observed by \cite{Atel7953} and this value may be related to the white dwarf spin period in the V5668 Sgr system, which is coincidental to the value of 71 s for the white dwarf spin period of DQ Her, e.g. \cite{kemp_circular_DQHer}. This nova type are generally associated with eruptions on the surface of CO white dwarfs and their maxima can be difficult to identify due to their jitter or oscillation features superimposed on an otherwise flat-top seen immediately prior to a distinguishable dust formation episode. Throughout this work, we refer to this early phase as the `flat-top-jitter' phase. The flat-top-jitter phase of the V5668 Sgr eruption was monitored by \cite{2017AN....338...91J}, where it was seen that the appearance of an increasing number of `nested P-Cygni profiles' in individual spectral lines could be associated with multiple ejection episodes or evolving components.

Several constraints of the system during the deepest part of the dust-dip are presented by \citet{banerjeeV5668sgr} from infrared high-cadence observations. Their observations resulted in a gas/dust temperature of $\approx$ 4000 K, a dust mass of $1\times10^{-8}$ M$_{\odot}$, and an expansion velocity of 530 km s$^{-1}$. \citet{banerjeeV5668sgr} found a black body diameter of the dust shell to be 42 mas on day 107.3. This estimate is sensitive to the fitted black body temperature of 850 K to the dust SED presented in Fig. 4 (right panel) of their work and corresponds to a physical diameter of $\approx$ $9.6\times10^{14}$ cm on the sky. 

Data is presented here from five nights of polarimetric observations acquired during the nova's permitted spectral phase with the Dipol-2 instrument mounted on the William Hershel telescope (WHT) and the La Palma KVA stellar telescope. The observations were obtained directly following the deep dust minimum and the nova's rise through its observed cusps. The nova shell of V5668 Sgr is not yet resolvable with medium-sized ground-based telescopes given the recent eruption, however we present a pseudo 3D photoionisation model based on 1D {\C} \citep{cloudy} models to demonstrate the ionisation structure of V5668 Sgr following its dust formation episode. 
Throughout the course of this work observations are mentioned in terms of days since discovery of the nova source and all quoted 
wavelengths are Ritz air wavelengths from the NIST database \citep{NIST}. 

\section{Observations}
\label{Observations}

\subsection{Polarimetry}\label{polarimetry}

\begin{figure}[ht!]
\includegraphics[width=9.5cm]{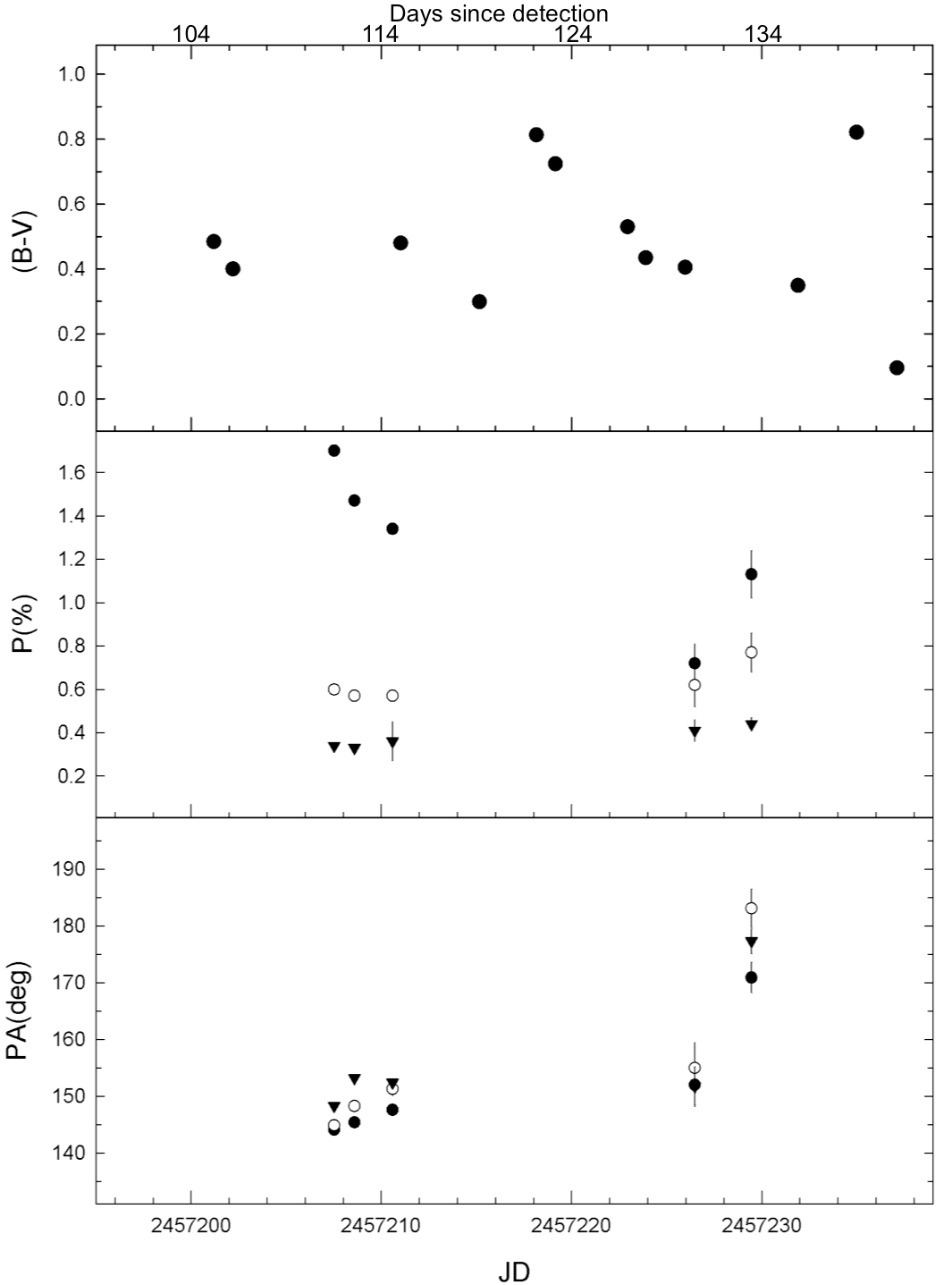}
\caption{Panels from top to bottom: (B-V) colour index as derived from AAVSO data during Dipol-2 observation epoch; the middle panel shows the measured absolute polarisation degree in percentage and the bottom panel shows the recorded position angles for the polarisation measurements. The filled circles shows the data for
the B-band, hollow circles - V-band and filled triangles - R-band. Days since outburst are marked along the top x-axis. The error bars ($\pm$1$\sigma$) are smaller
than the plotting symbol for the WHT data (days 111-114 post outburst).}
\label{fig:pol}
\end{figure}

The polarisation measurements of V5668 Sgr after the dust-dip stage were made
with the Dipol-2 polarimeter mounted on the 4.2m WHT telescope during three nights:
MJD2457207, 2457208 and 2457210 (days 111, 112 and 114 post-discovery). Two more measurements were recorded two weeks later with the 0.6m KVA stellar telescope (on MJD2457226 and 2457229, i.e. days 130 and 133 post-discovery, see Fig. \ref{fig:pol}).  Each night, 16 measurements were made of Stokes parameters q  and u and the weighted mean values computed. The exposure time was 10 sec for the WHT and 30 sec for the KVA. The
polarisation data, which have been acquired simultaneously in the standard B, V and R
pass-bands, are given in Table \ref{polarim}. 

Description of the polarimeter design is given by \citep{dipol2}.  Detailed descriptions
of the observational routine and data reduction procedure can be found in \cite{kosenkov}.

For determination of instrumental polarisation, we have observed a set of nearby ( d $<$ 30 pc)
zero-polarised standard stars. The magnitude of instrumental polarisation for Dipol-2 mounted
in Cassegrain focus on both telescopes was found to be less than 0.01$\%$ in all pass-bands, which is negligible in the present context. For determination of the zero-point of polarisation angle, we have observed highly polarised standards HD 161056 and HD 204827. The internal precision is 
{\raise.17ex\hbox{$\scriptstyle\sim$}}0.1$^{\circ}$, but since we rely on published values for the standards, the 
estimated uncertainty in determination of the zeropoint is less than 
1 - 2$^{\circ}$.

Effects that may be responsible for the observed variations in polarimetric measurements over the course of the observations include uncertainties regarding the standards, lunar proximity or orbital phase. Although, sky background polarisation is directly eliminated. Lunar proximity and seeing effects would add noise and contribute to larger errors rather than systematic deviations. 

\begin{table}

\caption{Polarimetry observations gathered during and rising out of the deep-dust dip experienced by this nova with the Dipol-2 instrument.}
 \centering
 \setlength\tabcolsep{2.5pt}
\begin{tabular}[width=0.5\linewidth]{lc c c c c cl}
     \toprule
        Date  (J.D.)     & Telescope   & 	Filter   & Pol (\%) $\pm$ err	& 	P.A. (deg) $\pm$ err\\
  \midrule
        2457207.5 & WHT & B & 1.699 $\pm$ 0.017 & 144.1 $\pm$ 0.3 \\
        2457207.5 & WHT & V & 0.601 $\pm$ 0.014 & 144.9 $\pm$ 0.7 \\
        2457207.5 & WHT & R & 0.344 $\pm$ 0.007 & 148.3 $\pm$ 0.6 \\
        2457208.5 & WHT & B &  1.471 $\pm$ 0.015 & 145.4 $\pm$0.3 \\
        2457208.5& WHT & V &  0.566 $\pm$ 0.012 & 148.3 $\pm$ 0.6 \\
        2457208.5 & WHT & R & 0.330 $\pm$ 0.006 & 153.2 $\pm$ 0.5 \\      
        2457210.6 & WHT & B &  1.338 $\pm$ 0.018 & 147.6 $\pm$ 0.4 \\
        2457210.6 & WHT & V & 0.565 $\pm$ 0.023 &  151.3 $\pm$ 1.2 \\
        2457210.6 & WHT & R & 0.357 $\pm$ 0.009 & 152.4 $\pm$ 0.7 \\
        2457226.5 & KVA & B & 0.723 $\pm$ 0.088 & 152.0 $\pm$ 3.5 \\
        2457226.5 & KVA & V & 0.616 $\pm$ 0.097 & 155.0 $\pm$ 4.5 \\
        2457226.5 & KVA & R & 0.416 $\pm$ 0.051 & 151.7 $\pm$ 3.5 \\
        2457229.5 & KVA & B & 1.132 $\pm$ 0.106 & 170.9 $\pm$ 2.7 \\
        2457229.5 & KVA & V &  0.770 $\pm$ 0.092 & 183.1 $\pm$ 3.4  \\
        2457229.5 & KVA & R & 0.440 $\pm$ 0.035 & 177.3 $\pm$ 2.2  \\
    \end{tabular}
\label{polarim}
\end{table}

\subsection{Photometry}\label{photometry}

Polarimetric observations were collected on 22 nights with the RINGO3 polarimeter \citep{ringo} on the LT \citep{Lpooltel} spanning days 113-186 after eruption detection. Unfortunately, it was found that the instrument's performance at low 
levels of absolute polarisation were not sufficient in the present context due to intrinsic non-negligible changes in the value of the EMGAIN parameter of the EMCCD detectors at the eight different positions of the polaroid rotor. This being said, the observations were of sufficient quality to perform differential photometry. The RINGO3 passbands were designed to incorporate the total average flux of a gamma-ray burst equally across the three bands and are thus unique to the instrument. The bands: known as red, green and blue, correspond to wavelength ranges 770-1000 nm, 650-760 nm and 350-640 nm respectively - roughly equivalent to the Johnson-Cousins I, R and B+V optical filter bands \citep{ringo}.

\begin{figure}[ht!]
\centering
\includegraphics[width=9.5cm]{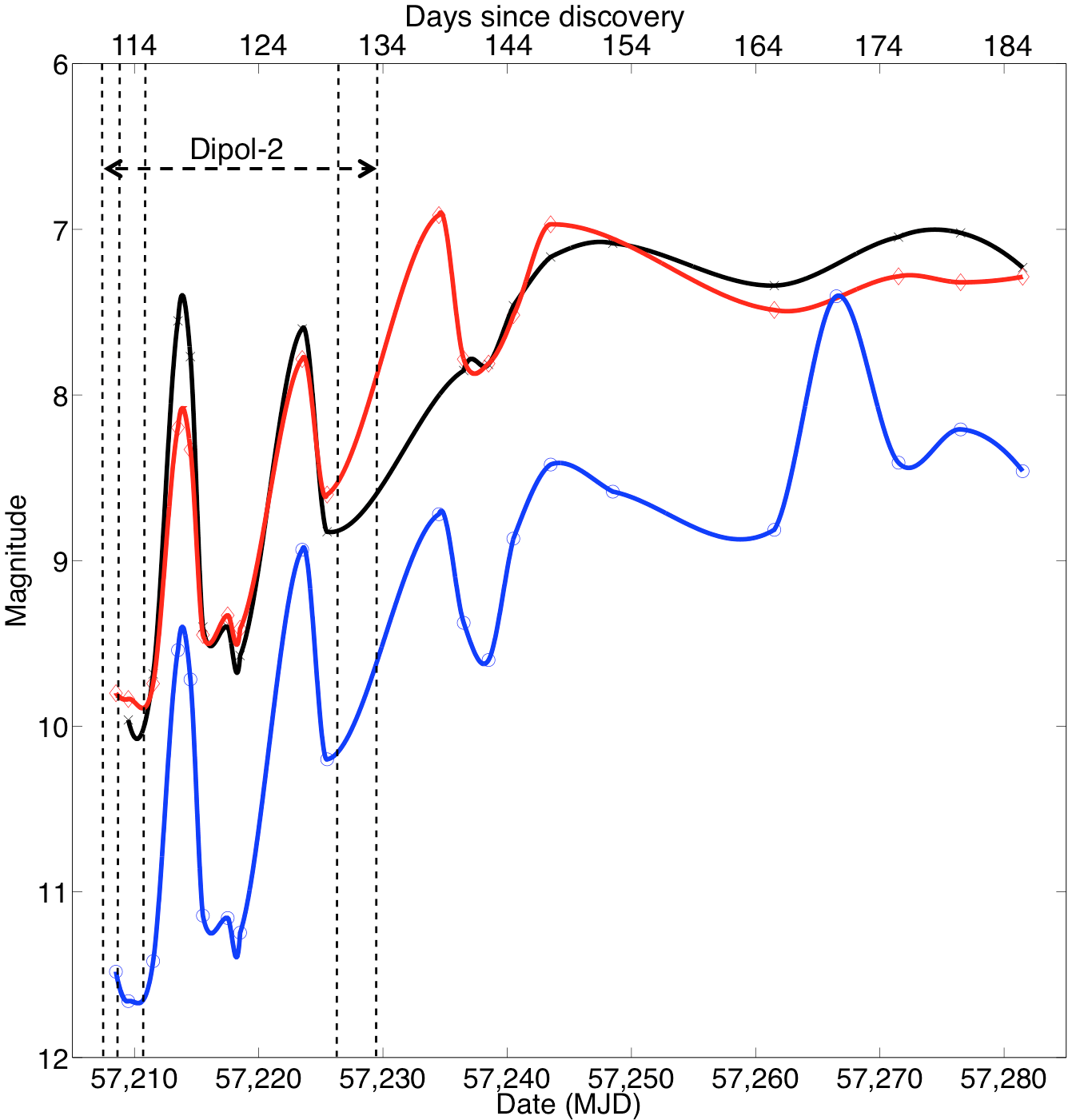}
\caption[V5668 Sgr differential photometry]{Results of differential photometry from the RINGO3 polarimetric data. The cusps on the rise out of the nova's 
dust-dip are clearly visible in the plot. The first cusp corresponds to the grain destruction seen in the Dipol-2 data days 
111 - 114 post-discovery. The rise on the third cusp feature visible in the plot is from the grain growth period suggested by the Dipol-2 observations, these observations lie between the dashed lines on the plot. The lines are colour-matched with the RINGO3 bands, blue is blue, red is red and green is black.}
\label{fig:polphot}
\end{figure}

The integrated flux from the 8 rotated exposures (S1) from each night of observation of the nova was recorded. The four brightest field stars were chosen for photometric comparison. The same field stars were not always within the frame on different dates. In essence, differential photometry was conducted with each one of the four field stars and the derived values were found to agree closely, in the end the brightest of the field stars was chosen as it gave the most reliable results and was present in the most frames across the different filters on the relevant nights. The results of this analysis can be seen in Fig. \ref{fig:polphot}. Performing photometry on this dataset allows for information to be gained on the nova systems behaviour during the sparsely populated AAVSO data during the Dipol-2 polarimetric as well as the densest auroral stage, during the FRODOSpec spectroscopic observation epochs.

\subsection{Spectroscopy}

Using FRODOSpec \citep{frodospec} mounted on the Liverpool Telescope  \citep{Lpooltel}  in low resolution mode, spectra were taken over 
103 nights from outburst detection until day 822 post outburst. Data acquired with FRODOSpec are reduced and 
wavelength calibrated through the appropriate pipeline, detailed by \cite{frodospec}. The spectra were flux 
calibrated using standard routines in \textsc{iraf} \footnote{{\sc iraf} is distributed by the National 
Optical Astronomy Observatories, which are operated by the Association of Universities 
for Research in Astronomy, Inc., under cooperative agreement with the National Science Foundation.} 
\citep{todyiraf} against a spectrum of G191-B2B taken on 30 Sept 2015 using the same instrument setup. 
The standard spectral data was obtained from \cite{oke}. All spectra were taken in the low-resolution mode 
of FRODOSpec except for two dates, these being day 411 and 822 post-discovery, whose observations were 
collected in high-resolution mode. The resolving power of the low-resolution mode are 2600 for the blue 
arm and 2200 for the red arm. The low-resolution mode in the blue arm therefore gives a resolution of around 
1.8 $\rm{\AA}$ or 120 km s$^{-1}$. High resolution mode has a resolving power of 5500 in the blue arm in 5300 
for the red arm. The spectroscopic data were analysed using SPLOT and other standard routines in {\sc iraf}. 
As the vast majority of the spectra involved are from the low resolution mode a systematic error of up to 
20\% is expected as well as a 10\% random error.


\section{Analysis and Results}

\subsection{Polarimetry}
\label{pol_anyl}

The wavelength dependence of polarisation (sharp increase towards the blue) gives
strong support for Rayleigh scattering as the primary source for the observed polarisation
after the dust formation stage; see \cite{Atel7643,Atel7862} for more on this particular nova's 
primary dust formation episode. The directions of polarisation in the B, V and R bands
are close to each other for the five dates, which suggests an intrinsic nature to the
observed polarisation. The interstellar component is small even in the V and R bands
because the angle of polarisation in the V and R bands are always close to that in the B-band.

Unfortunately, photometry data at the dates when the polarisation with
the Dipol-2 instrument was measured is not available. 
As can be seen from Fig. \ref{fig:pol}, however, the 
color index (B-V) did not change significantly over the range of dates from when the absolute polarisation 
was measured. Consequently, the rapid changes in the B-band polarisation seen on days 111-114 and 130-133 
post-discovery are not due to decrease or increase of the fraction of the polarised scattered light in the system.

Variations in the absolute polarimetry over the five nights observing of V5668 Sgr with Dipol-2 
covered days 111-133 post-discovery are after the formation of dust and during the 
local minimum in the transitional-stage of the optical light curve. The observed flux in the 
R-band is likely dominated by H$\alpha$ in these observations. The most probable 
explanation for the observed variations in the absolute polarisation is the small
dust particles which are responsible for the appearance of the polarised scattered light.
During days 111-114 post-discovery, the destruction phase could be observed
while during 130-133 post-discovery, the creation phase was recorded. 

X-ray counts increased during the observations reported here, see \cite{Atel7953}, Gerhrz et al. 
(in prep.), exposing the nova shell to a harsher radiation field. Infrared SOFIA observations 
\citep{Atel7862}, that coincided with the commencement of the observations presented here, 
revealed that the dust emission on day 114 post-discovery had increased since day 83 post 
discovery and that a reduction in grain temperature suggested rapid grain growth to sub-micron radii. 
These observations suggest that the hydrogen emission in the NIR was blanketed by the dust and 
the effect of this can be seen in the strengthening of the Paschen series, see Figs. \ref{fig:spec1} 
\& \ref{fig:spec2}. Emission at this time is expected to arise from a cold dense shell as well as hotter, 
less-dense ejecta, see \cite{derdzinski}. Gehrz et al. (in prep) are presenting $\it{Swift}$ and 
SOFIA observations of this nova covering the IR, UV and X-ray behaviour of the nova system. 

UBVRI polarimetry before the dust-dip, taken during the flat-top-jitter phase over the first 
observed major primary jitter, was reported by \cite{muneer}, providing knowledge of the absolute 
polarisation of the system before the major dust formation event. Although the measurements by 
\cite{muneer} are not corrected for interstellar polarisation toward the source, the correction 
is not made here either. The earlier \cite{muneer} results are lower than those of the Dipol-2 
observations directly following the dust-dip, with position angle (P.A.) measurements being consistent. 
The earlier observations with lower recorded absolute polarisation of \cite{muneer} fit 
with electron scattering or interstellar polarisation, as expected before the dust formation episode. 
Of interest regarding the observations of 
\cite{muneer} is that between days 2 - 4 post-discovery, P.A. of the 
polarisation varies between roughly 150$\rm{^{o}}$ and 10$\rm{^{o}}$ (i.e. 190$\rm{^{o}}$) - 
very similar to that in the Dipol-2 measurements, see Table \ref{polarim}. In the observation presented here, 
values between 144$\rm{^{o}}$ - 183$\rm{^{o}}$ were found for the P.A. 
The origin location of the source of the polarisation is indicative of the opening angle of the component, 
be it the equatorial or polar nova shell components is unknown. The work of \cite{derdzinski} would suggest the opening 
angle to be related to the equatorial disk.

These observations can be understood in terms of dust resulting from seed nuclei that formed 
during the optical dust-dip and increase of the density in the forward shock zone. \cite{lynch_cusps,kato2009} 
and \cite{Strope:2010aa} discuss cusps as possibly arising from shocks in the nova outflow. Since V5668 Sgr 
is a slow nova, strong shaping of the ejected nova material is expected. A strong correlation in position 
angle of the polarisation is needed throughout the observed epochs if it is related to either the equatorial 
or polar components of the nova outflow. 
The shock passes through the layer of fresh-formed small dust grains and destroys them, yet retaining seed 
nuclei, thus allowing for the process to repeat over the next shock cycle. $\it{Swift}$ X-ray data 
(Gehrz et al. 2017, in prep) shows the X-ray count rising on entering the dust-dip and increasing again 
when the cusps start (post dust-dip-minimum). The phenomenology of the hard X-rays can be understood in 
the context of shocks and sweeping up material, allowing the local densities to increase, which creates 
favourable conditions for dust formation. The soft component of the X-ray emission should be due to 
continued nuclear burning on the surface of the central white dwarf \citep{Landi:2008aa}, whereas the hard component is 
expected to arise from shocks, e.g. \cite{metzger}.
 
Gamma-ray emission was observed for the V5668 Sgr nova event and is described by \cite{cheung_gammav5668sgr}. 
The emission of gamma-ray photons of energy $\geqslant$100 MeV lasted around 55 days, longer and intrinsically 
fainter than any of the six other nova observed to produce gamma-ray emission. The onset of gamma-rays occurred 
two days following the first optical peak and were followed for 212 days. Due to low photon counts the team 
who discovered the sixth confirmed gamma-ray nova, were unable to correlate gamma variability with that in 
the optical, although the gamma emission peaks during the third major jitter (around days 30 - 40 post-discovery) 
on the nova's otherwise flat-top light curve. The Fermi-LAT observations of this nova ended one month previous 
to the observations discussed here and before V5668 Sgr's dust formation event.


\subsection{Spectroscopy}
Observed spectra were calibrated and subsequently interpreted using published results from the 
literature and {\C} simulations \citep{cloudy}. Shocks suggested by the 
polarimetry and multi-wavelength observations discussed in \ref{pol_anyl}. 

The earliest spectra observed here are interesting from the point of view of a suggestion of 
multiple components during the flat-top-jitter phase. A spectrum obtained on day 0 post-discovery 
shows P-cygni profiles with absorption components at -1200 km s$^{-1}$. Further observations 14 days 
later revealed two absorption components in the strongest spectral lines with each having a 
measured velocity of -950 and -520 km s$^{-1}$ in the Balmer lines. In spectra taken in mid April, 
the highest velocity component of the maximum spectrum increased again to approximately -1125 km s$^{-1}$ 
and with a new lower-velocity component of -610 km s$^{-1}$. These observations hint at optical 
depth effects where in the 14 days post-discovery spectrum, the inner side of the expanding shell is visible. 
Then by day 27, an outer shell section may becomes visible when three absorption components appear with 
velocities of -554, -945 and -1239 km s$^{-1}$, respectively. The expanding shell is still 
expected to be radiation bound at this stage due to the high densities present. On day 28, the observed velocities decrease 
to -507, -887 and -1065 km s$^{-1}$. The next spectrum was observed on day 31 with FRODOSpec, where 
it can be seen that the middle absorption component disappeared and leaving two components at -537 and 
-1047 km s$^{-1}$. In spectra taken on days 32 and 33 post-discovery, a slight increase is 
seen in the absorption components which then levels off until the absorption systems disappear and are 
replaced by emission wings. In the subsequent spectra it appears that only the slower component 
remains visible as part of the expanding shell. It is worthy to note that the appearance of additional 
absorption components appear to be correlated with the local maxima in the nova's early light curve. 
The evolution of the aforementioned absorption components can be seen in Figs. \ref{fig:earlyspecqueue} 
\& \ref{fig:earlyspec12wit}. The Ca~{\sc ii} lines during these early days display a 
similar structure to Balmer and nebular [O~{\sc iii}] lines at late times. 

The spectra presented in Figs. \ref{fig:spec1} and \ref{fig:spec2} show 10 nights: from 6 July to 14 August. 
It was found that the earlier July spectra (days 114, 116, 120, 122 and 123 post-discovery), see Fig. \ref{fig:spec1}, 
are all quite similar in appearance and it can be noted how the observed flux from the system recovers. 
These five spectra correspond to the early-rise out of the dust-dip while the shell is known to be mostly 
optically thick, exhibited by the presence of strong permitted lines. 

\begin{landscape}
 \begin{figure*}[ht!]
\centering
\includegraphics[width=15cm]{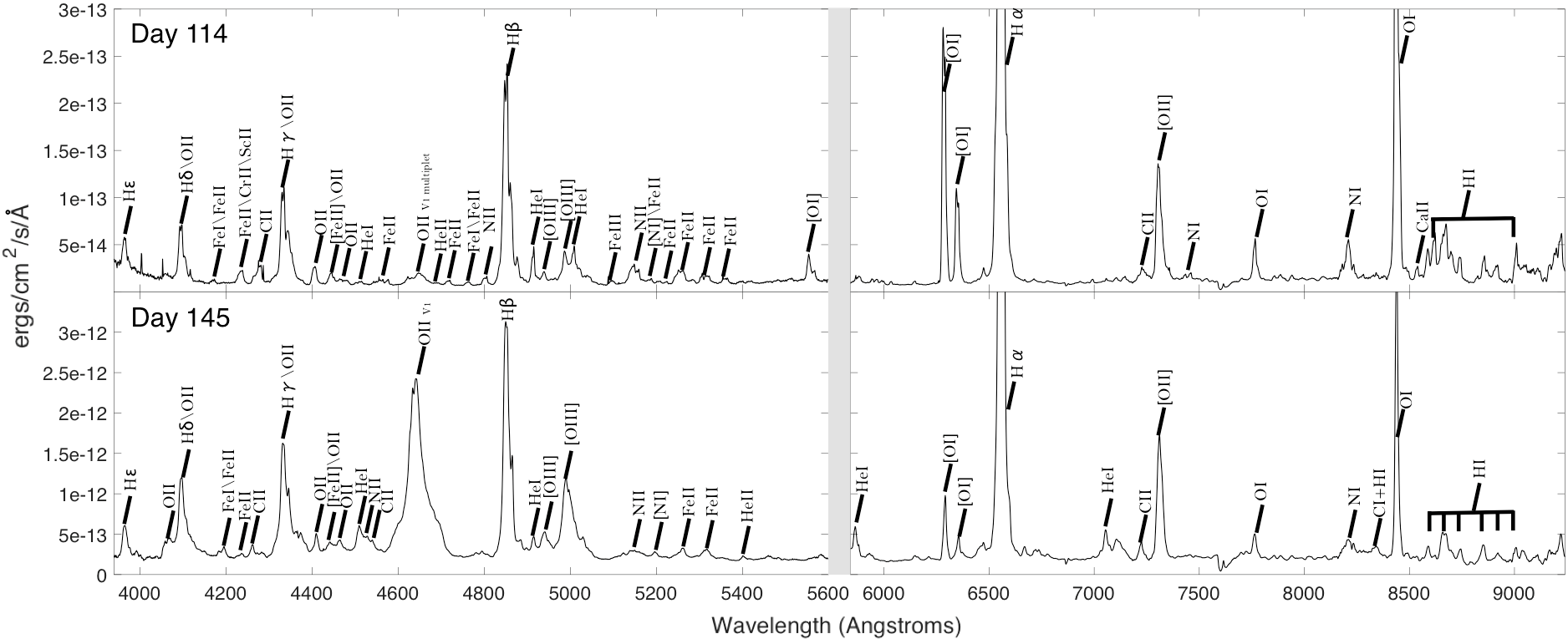}
\caption[V5668 Sgr spectra 1]{Spectra showing the two main spectroscopic stages observed during the polarimetric observations. Note the fall in O~{\sc i} corresponding to a rise in O~{\sc ii} and O~{\sc iii} lines.}
\label{fig:spec3}
\end{figure*}

\begin{figure*}[ht!]
\centering
\includegraphics[width=15cm]{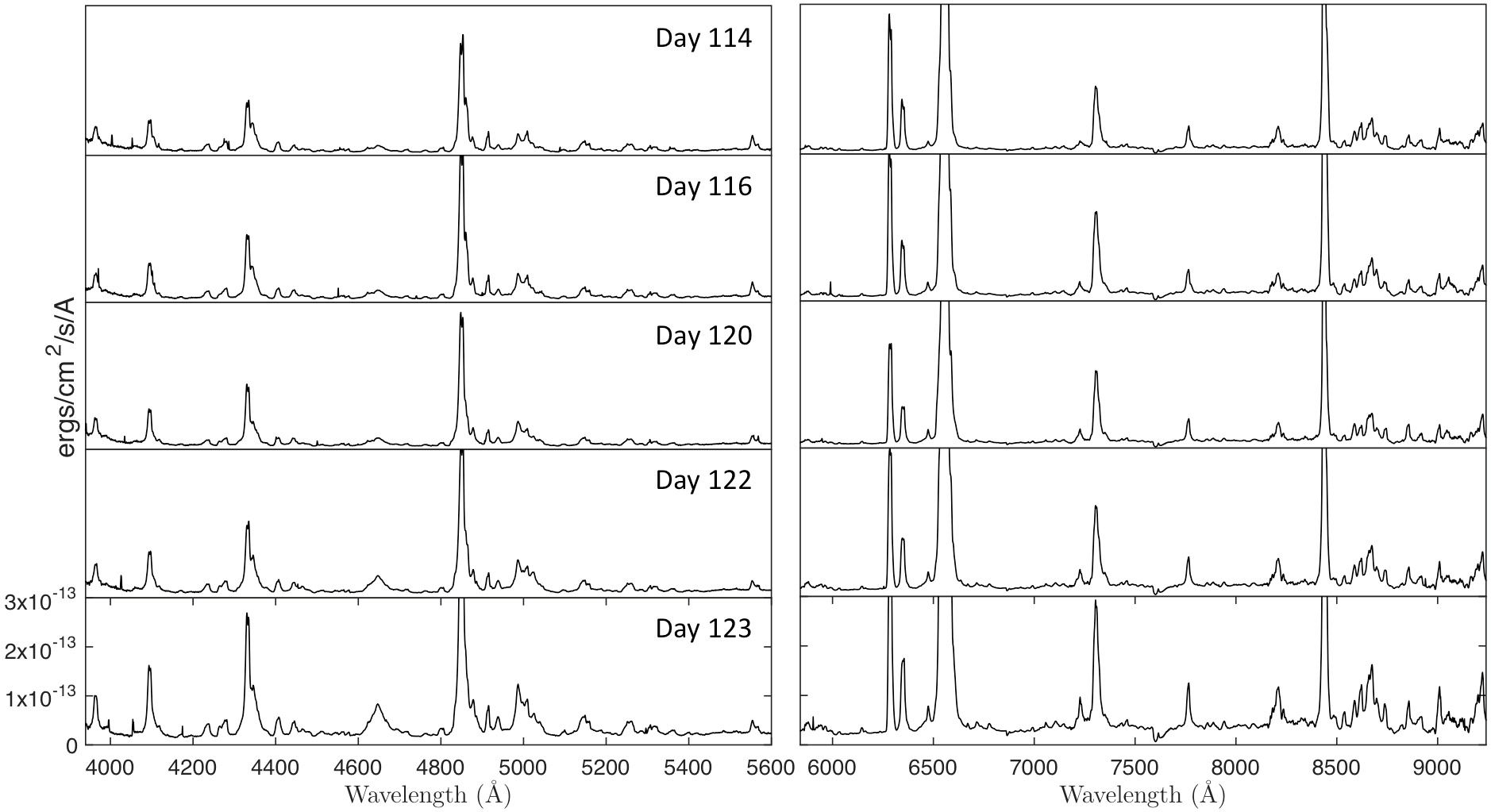}
\caption[V5668 Sgr spectra 2]{Early July 2015 spectra from FRODOSpec with the blue arm spectra in the left hand series and red arm on the right. 
Spectra were flux calibrated by Dr. Steve Williams, and are all scaled to the bottom panel of each column. Note the change in flux over the dates.}
\label{fig:spec1}
\end{figure*}

\begin{figure*}[ht!]
\centering
\includegraphics[width=15cm]{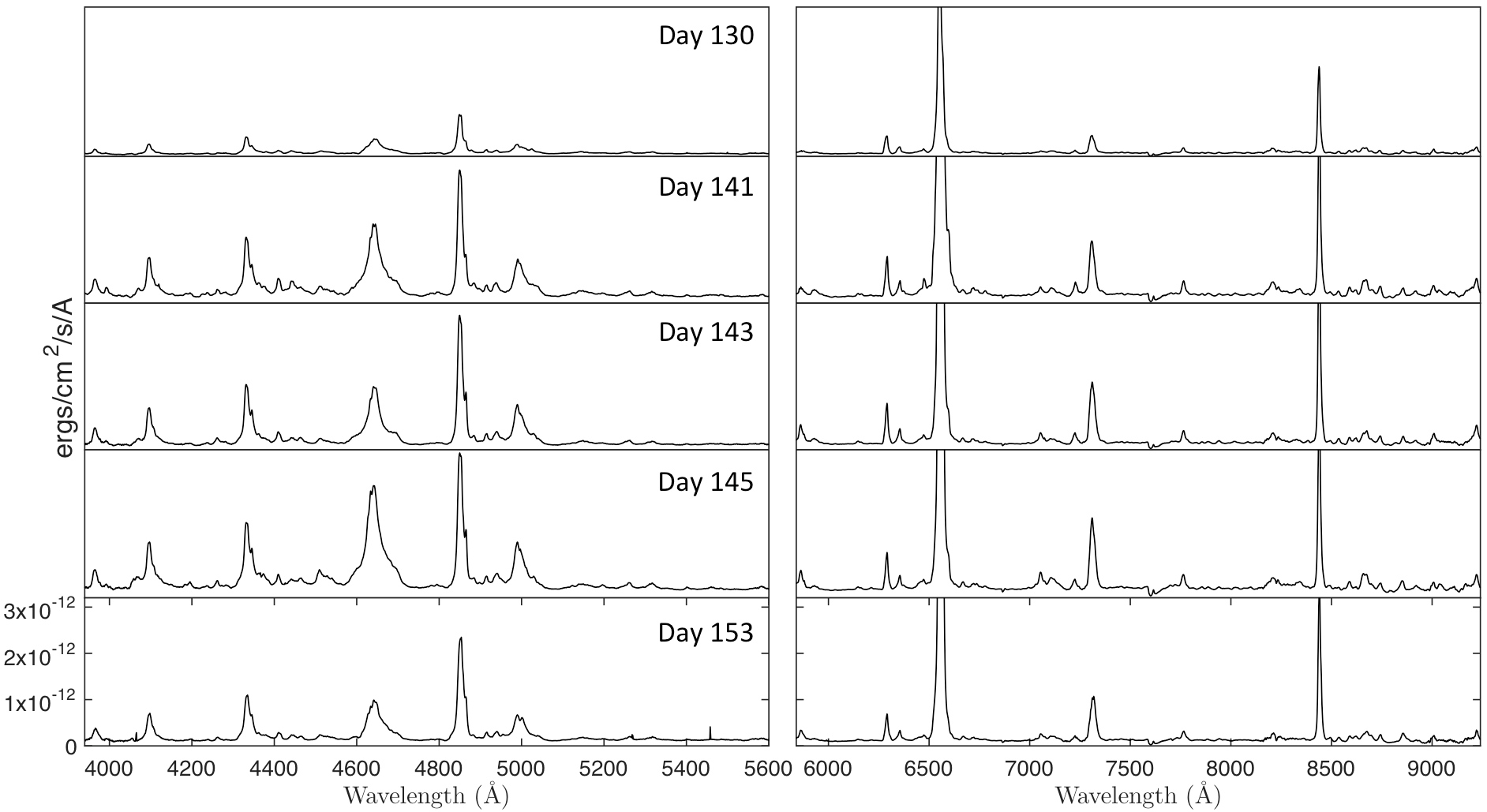}
\caption[V5668 Sgr spectra 3]{Same as in Fig. \ref{fig:spec1} except for spectra taken in late July and early August 2015, using FRODOSpec. 
Note the flaring feature in the blue arm of the spectra identified in this work as arising from the O~{\sc ii} V1 multiplet, 
see Section \ref{oxfla}.}
\label{fig:spec2}
\end{figure*}
\end{landscape}
 
\begin{figure}[ht!]
\centering
\includegraphics[width=8cm]{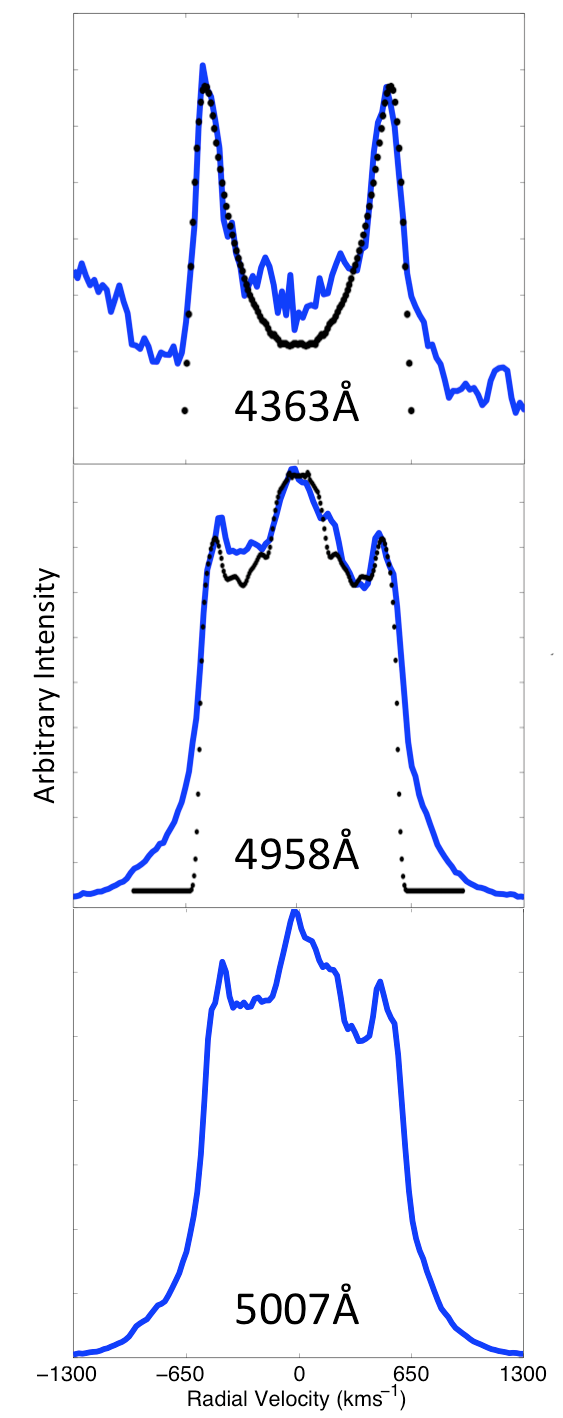}
\caption[V5668 Sgr spectra 5]{Spectra of [O~{\sc iii}] nebular and auroral lines on day 822 post-discovery. The observed 
line profiles (blue-solid lines) were used in the fitting of a morpho-kinematical model with the {\Sh} software, seen as 
the overlaid black dots. The auroral line is fitted with an equatorial disk whereas the nebular lines fit an equatorial 
waist and polar cones morphology with a Hubble outflow velocity law, see Fig. \ref{fig:shape}. }
\label{fig:spec4}
\end{figure}

\begin{figure}[ht!]
\centering
\includegraphics[width=9.5cm]{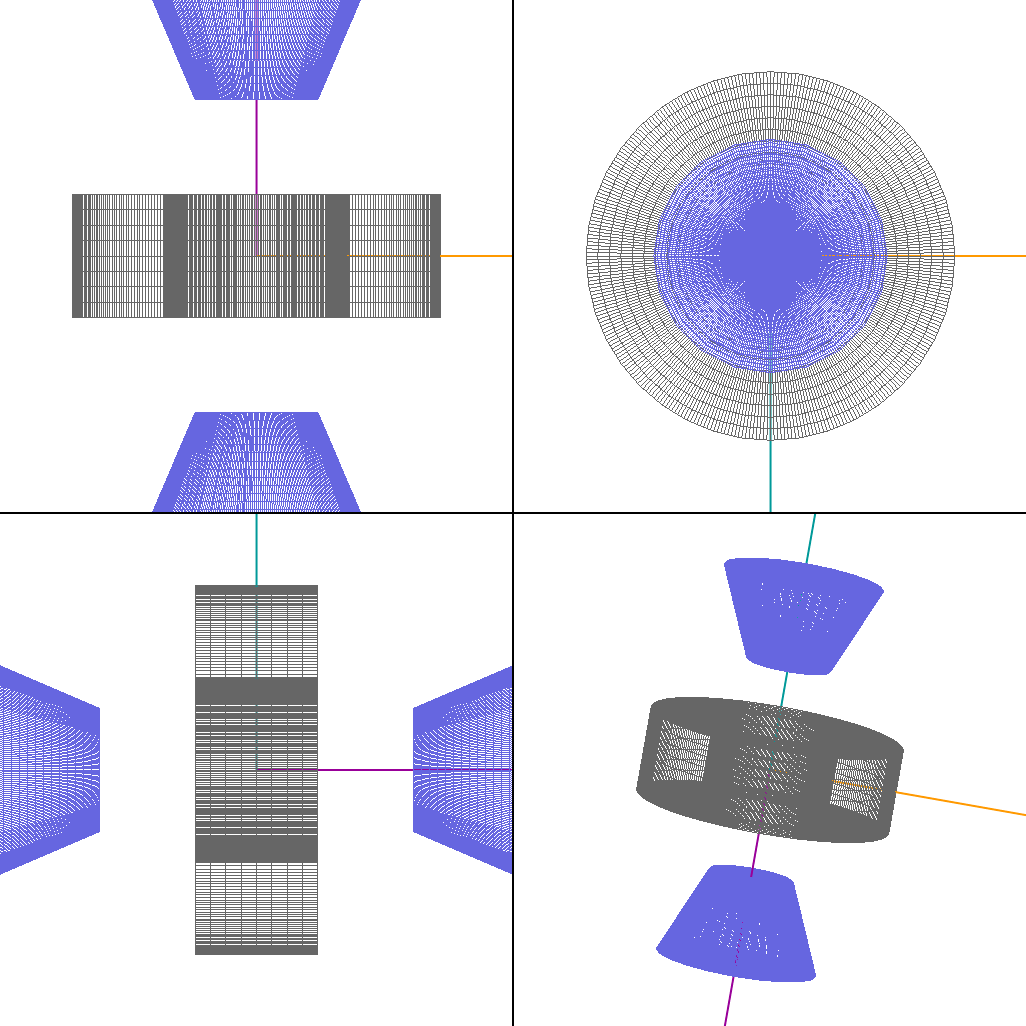}
\caption[V5668 Sgr {\Sh} model]{Mesh display of the spatial structure of the nova shell around V5668 Sgr as determined from the nebular [O~{\sc iii}] line shapes with 
the {\Sh} software, see Fig. \ref{fig:spec4}. The four panels show different orientations of the structure, the bottom-right-hand-panel shows the nova 
shell placed at the P.A. suggested from the polarimetry and as visualised in the py{\C} renderings seen in Fig. 
\ref{fig:pycloudy}. If the detected polarisation has its origin in the equatorial waist then the shell should be titled at 
90$\rm{^{o}}$ in the plane of the sky. The saddle shaped 4363 $\rm{\AA}$ [O~{\sc iii}] line feature, as well as other higher 
excitation species, can be understood as arising from the equatorial-ring-waist.}
\label{fig:shape}
\end{figure}

During the optical light curve's ascension out of the 
dust-dip, the most interesting changes are observed in the spectra from 130, 141, 143, 145 and 153 post 
discovery (see Fig. \ref{fig:spec2}). The final three spectra presented from days 143, 145 and 153 following 
the eruption straddle a major cusp on the way out of the nova's visual dust-dip, see Fig. \ref{fig:polphot}. 
As discussed in Section \ref{pol_anyl}, these cusps are commonly associated with shocks that occur in 
the immediate aftermath of the eruption. The most striking feature has been referred to as `nitrogen flaring' 
in many previous works, see e.g. \cite{tololo,zemko}, around 4650 $\rm{\AA}$.

Over the same time frame, Ca~{\sc ii} is observed to decline whilst He~{\sc i} and He~{\sc ii} are both 
observed to increase. Fe~{\sc iii} and N~{\sc i} emission strength decrease along with the Paschen 
series with respect to H$\beta$; for more details see Table \ref{fluxl}. The observed behaviour is consistent with 
the thinning out of ejecta, which subjects the gas mix to harder radiation from the central source.

\subsection{Simulations}
\label{cloudys}

The spectral development of V5668 Sgr is dominated by the Balmer series plus He, N, O and Fe lines  as the nova progresses through 
its permitted, auroral then nebular spectral stages. According to the Tololo classification scheme 
the nova is in its P$\rm{^o_o}$ stage during the observations presented in this work. A parameter sweep was conducted using 
the python wrapper for {\C} \citep{cloudy} known as py{\C} \citep{pycludy} to examine the line ratios for the hot-dense-thick 
nova shell that is still close to the burning central system. It was found that the dust shell size of \cite{banerjeeV5668sgr}, 
when extrapolated to the expected size for the dates under study in this work, can fit the observed spectra although better 
fits can be achieved with marginally smaller radii, hinting that the optical emission lines in the optically thick region 
inner to the dust shell. 
An implication is that dust clumps should appear and then disappear along the line of sight to the observer, further complicating 
the analysis. 

Initial parameter sweeps were coarse and broad covering densities of $10^{4} - 10^{14}$ cm$^{-3}$. It was found that densities from 
$10^{8} - 10^{10}$ cm$^{-3}$ better 
explained the structure of the observed spectrum and refined grids were run over these constraints. 
It is cautioned that, at the densities studied here, the \cite{nussbaumerstorey} CNO recombination coefficients 
used are not as reliable since the LS coupling scaling law assumed in \cite{nussbaumerstorey} diverge for atoms 
with upwards of two valence electrons.

Fig. \ref{fig:cloudy1} shows the results of a parameter sweep 
including log densities 8.60 - 9.20 in 0.05 dex, and the effective temperature of the central source 
from $6\times10^{4} - 3.0\times10^{5}$ K in steps of $2\times10^{5}$ K. 
For the parameter sweep py{\C} was 
used to control {\C}. An average of Fe~{\sc ii} type nova abundances adapted from \cite{warner} were included. The Eddington 
luminosity of a 0.7 M$_{\odot}$ white dwarf was assumed, with r$_{min}=3.2\times10^{14}$ cm and r$_{max}=6.4\times10^{14}$ cm. 
As the binary characteristics of this system are not known, a 0.7 M$_{\odot}$ white dwarf was chosen based on the turn-on time 
in X-rays (Gehrz et al. 2017, in prep.) and the nova's t$_{2}$ value in comparison to Fig. 4(c) of \cite{henze}. From 
this type of analysis, it is possible only to say that the white dwarf must be on the lower end of the scale found in nova progenitor 
systems and that it is most probably a CO white dwarf.

\begin{figure}[ht!]
\centering
\includegraphics[width=10cm]{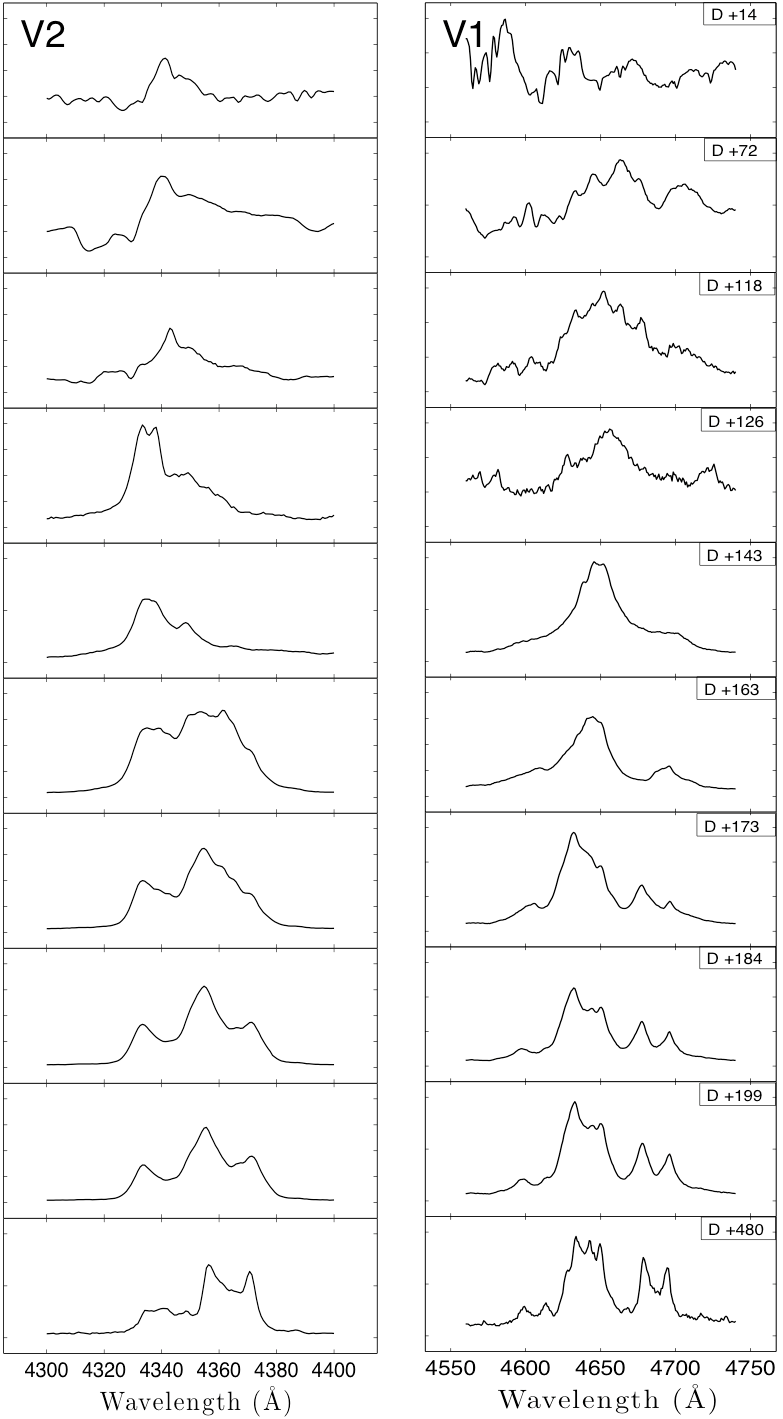}
\caption[Line shape evolution in V1 and V2 multiplets]{Temporal shape evolution of the blending lines in the area surrounding the 
4341 $\rm{\AA}$ and 4650 $\rm{\AA}$ V2 and V1 oxygen 
multiplets. Days since detection are marked in the top-right-corner of each subplot in the V1 column. The most pronounced 
flaring episodes are between days 140 - 150 post-discovery. Note in the V1 multiplet column plot (right hand side) a saddle-shaped 
He~{\sc ii} line at 4686 $\rm{\AA}$ fits the 4676 and 4696 $\rm{\AA}$ lines if they are the red and blue wings of the He~{\sc ii} line.}
\label{fig:spec5}
\end{figure}

\begin{landscape}

\begin{figure}[ht!]%
    \centering
    \subfloat[][]{{\includegraphics[width=13cm]
    {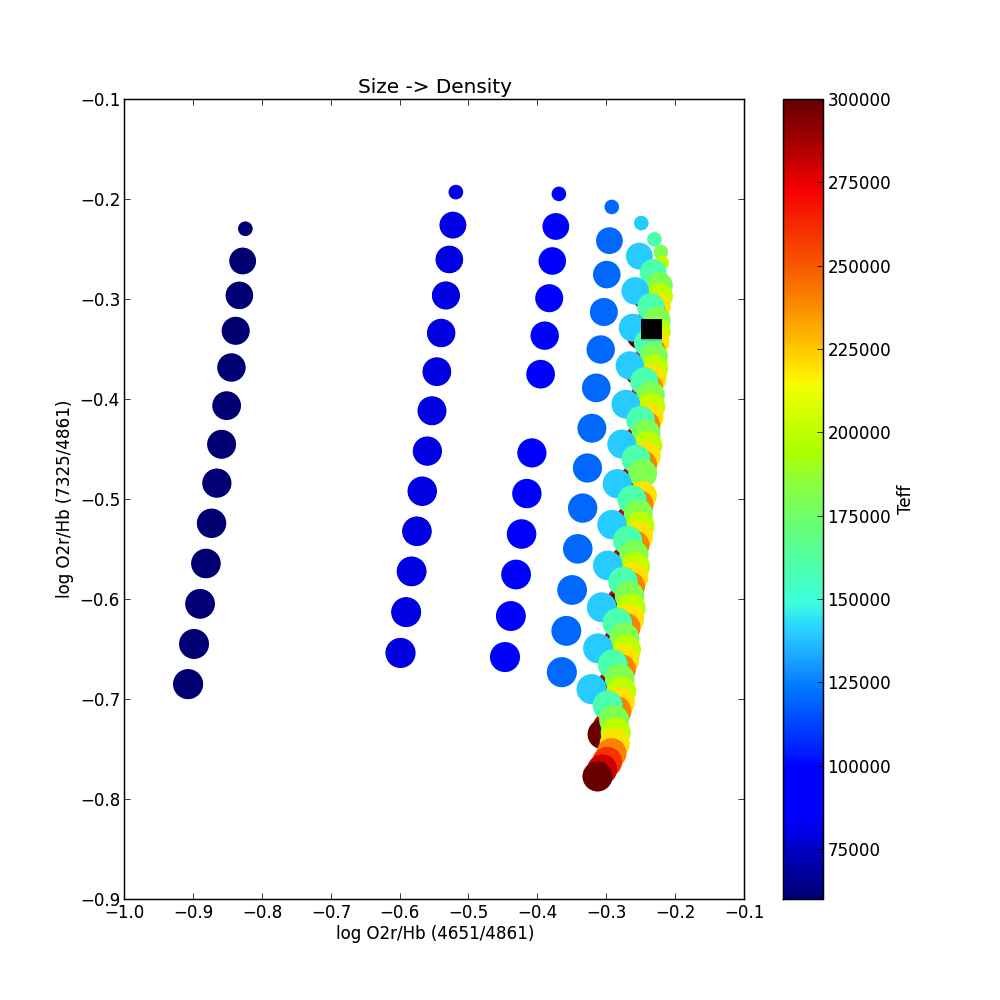} }}%
    \subfloat[][]{{\includegraphics[width=13cm]{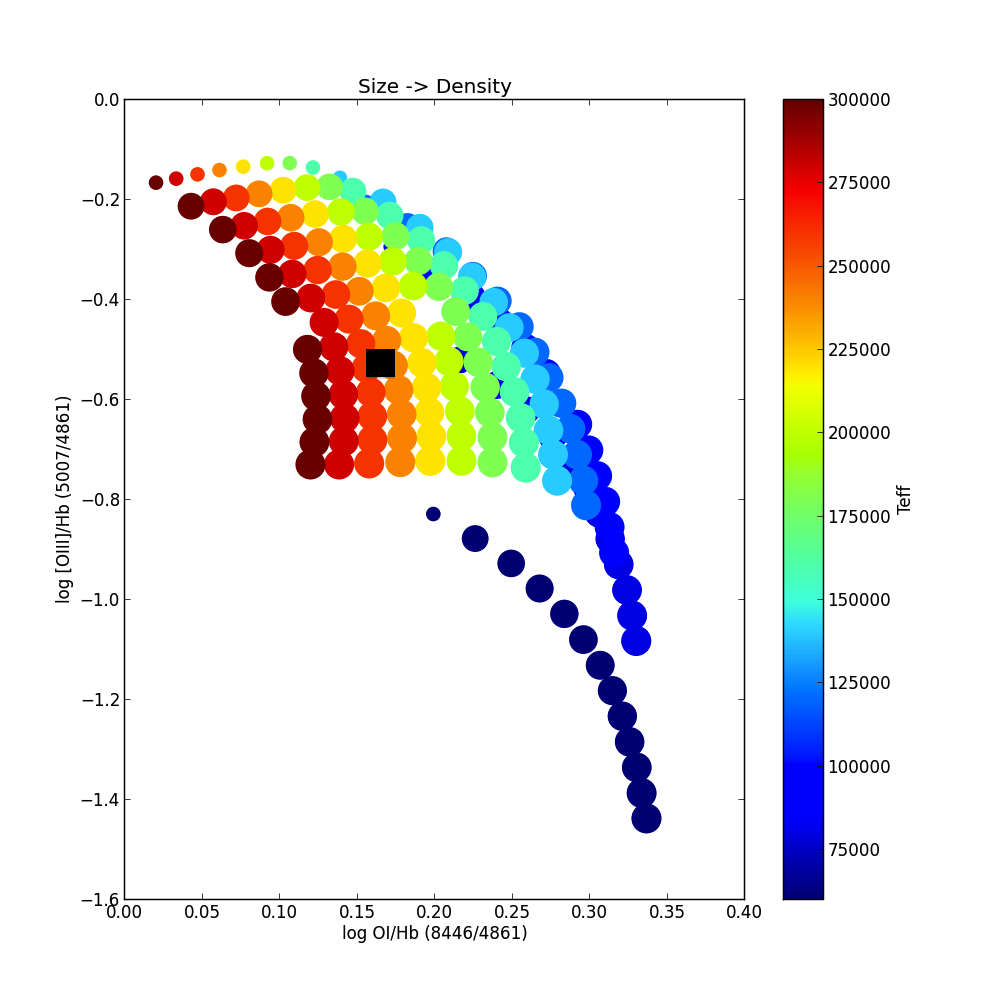} }}%
    \caption{ 1D {\C} parameter sweep of temperature and density of two O~{\sc ii} multiplets relative to H$\beta$. 
    (a) shows O~{\sc ii} 4650 $\rm{\AA}$/H$\beta$ and 7325 $\rm{\AA}$/H$\beta$ diagnostic diagram, 
    whereas (b) shows a {[O~{\sc i}]} 8446 $\rm{\AA}$/H$\beta$ and {[O~{\sc iii}]} 5007 $\rm{\AA}$/H$\beta$ diagnostic diagram. Displayed here 
is a zoomed in version of the parameter sweep covering 
the log of the nebular densities vary from 8.60 - 9.20 in 0.05 dex, and the effective temperature of the central source 
from $6\times10^{4} - 3.0\times10^{5}$ K in steps of $2\times10^{4}$ K. The black squares mark the measured rations on day 141. }%
\label{fig:cloudy1}
\end{figure}

\end{landscape}

\begin{figure*}[ht!]
\centering
\includegraphics[width=15cm]{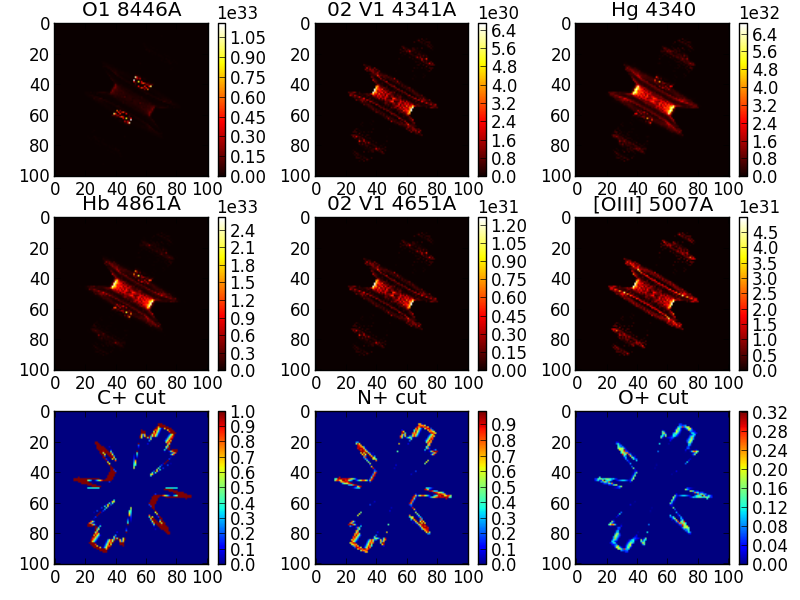}
\caption[V5668 Sgr photoionisation model]{{\textsc{pyCloudy}} emission model of V5668 Sgr, using nova abundances with the inclination angle set at 
85$^{\circ}$. Conditions for the model were the best derived values from the parameter sweeps assuming luminosity and geometries discussed in Sect. \ref{cloudys} for day 141 after discovery. 
The input {\Sh} model is seen in Fig. \ref{fig:shape}. Volumetric flux units are in ergs s$^{-1}$ (colour bars). The \textit{x} and \textit{y} axis values are scaled units 
of physical size, r$_{min}$ and r$_{max}$ determined from \cite{banerjeeV5668sgr} size on day 107 post-discovery. Clumpiness was 
simulated using a Perlin noise modifier in {\Sh}.}
\label{fig:pycloudy}
\end{figure*}

The best fitting densities from the {\C} parameter sweep were in the range $6.3\times10^{8}$ - $1.0\times10^{9}$ cm$^{-3}$ and an 
effective temperature of $(1.8 - 2.4)\times10^{5}$ K was found for the chosen radial distance, luminosity and abundances for day 141 post 
discovery. With 
information from the polarimetry and spectroscopy on conditions present in the expanding nova shell, an attempt to visualise 
the unresolved shell is presented in Fig. \ref{fig:pycloudy}, where the models are valid for day 141 post-discovery. 
In the top six panels of Fig. \ref{fig:pycloudy}, are the simulated emission of oft-seen oxygen emission lines in erupting 
nova systems. A comparison of the locality of emission through the shell of the same species is presented in each column 
of each of these panels. The 
O~{\sc i} line the strong emission produced from the simulated 6 level oxygen atom line of 8446 $\rm{\AA}$ at the inside of the shell 
is in good accordance with observations, see also Fig. \ref{fig:spec5}(b). The O~{\sc ii} panels (middle column Fig. \ref{fig:pycloudy}) simulate both the V1 and V2 
multiplets that are shown relative to H$\beta$ in Fig. \ref{fig:spec5}(a) and discussed in \ref{oxfla} of this work. 
The [O~{\sc iii}] panel 
demonstrate the locality and relative strength of the nebular 5007 $\rm{\AA}$ line.  
The bottom three panels are, from left to right, ionic cuts of C, N and O, respectively. The {\Sh} \citep{shape} line 
profile model 
fits are to day 822 post-discovery, by when the line structure had frozen, and can be seen in Figs. \ref{fig:spec4} \& 
\ref{fig:shape}. A Perlin noise modifier was applied to the hydrogen density distribution of the polar cones and equatorial ring, 
with the average density being $1.0\times10^{9}$ cm$^{-3}$. The 
luminosity was set to log(L$_{\odot}$) = 4.36, and an effective temperature of $1.8\times10^{5}$ K was assumed based 
on the parameter sweep, see Fig. \ref{fig:cloudy1}. To simulate the nova conditions on day 141 post-discovery an inner radius of $3.2\times10^{14}$ cm 
and an outer radius of $6.4\times10^{14}$ cm were assumed, as in the parameter sweep.



In order to create the {\Sh} model, subsequently used for input into py{\C}, three components were used comprising the 
equatorial waist and the two polar cones, see Fig. \ref{fig:shape}. The ring-like waist was constructed from a 
cylinder primitive in which a density, Hubble velocity law and thickness were applied. The two polar 
features were constructed using cone primitives. The densities applied to the features were estimated 
using {\C} simulations and the velocity components were found from measuring Doppler broadening of the 
Balmer emission lines. Emission line structure in fast outflows depend strongly on their velocity field 
and orientation, {\Sh} allows the user to untangle the projection effects. The frozen line shapes of the 
nebular stage modelled in Fig. \ref{fig:spec4} are with a inclination of 85$^{\circ}$ a polar velocity of 
940 km s$^{-1}$ and equatorial velocity of 650 km s$^{-1}$ at their maximum extensions. The proposed 
structure is similar to that found in other slow novae such as T Aur (see Section ) and DQ Her. 

\subsection{Oxygen Flaring}
\label{oxfla}

It is well known that identification of emission lines in nova eruption spectra is difficult, largely due to blending and large Doppler widths. Table \ref{V1V2} and Fig. \ref{fig:spec5} demonstrate two cases where important diagnostic lines can easily be confused for other lines or are heavily blended 
without realisation. Through simple additive arguments based on A$_{ki}$ values it was found that the O~{\sc ii} V 1 multiplet is 
comparable to the commonly identified N~{\sc iii} and C~{\sc iii} lines. 
As there is mention but no modern discussion on O~{\sc ii} 
in the place of this `flaring' feature, its nature was investigated. Collisional rates are of importance but are not well constrained.

Focusing on the five spectra presented in Fig. \ref{fig:spec2}, we observe the `flaring' episode around 
4650 $\rm{\AA}$. This type of flaring episode is often attributed to `nitrogen flaring', although in the photoionisation 
simulations (Fig. \ref{fig:cloudy1}) the lines under derived conditions do not favour pumping of N~{\sc iii} lines through the Bowen fluorescence 
mechanism nor the ionisation and subsequent recombination of the C~{\sc iii} lines. Instead, the recombination 
of O~{\sc ii} around 4650 $\rm{\AA}$ appears responsible for the majority of the emission, with some contribution expected 
from pumping of the same and contribution from the Fe~{\sc iii} 4658 $\rm{\AA}$ line. The He~{\sc ii} line at 4686 $\rm{\AA}$ 
may contribute to the red end of the observed blend, which can be seen in Table \ref{V1V2} and Fig. \ref{fig:spec3}, He~{\sc ii} 
4686 $\rm{\AA}$ in a saddle-shaped line profile with emission components around $\pm$ 520 km s$^{-1}$ would appear the 
same as the two longest wavelength lines in the O~{\sc ii} multiplet in the region, i.e. at 4676 and 4696 $\rm{\AA}$. 
As a consequence of this the presence of He in a nova cannot be confirmed with only the presence of the He~{\sc ii} 
4686 $\rm{\AA}$ emission line.

The concept of nitrogen flaring dates back to 1920 when \cite{fowler} identified an `abnormal' strong spectral feature peaking around 4640 - 4650 $\rm{\AA}$. Following this, Mr. Baxandall and W. H. Wright exchanged letters regarding Prof. Fowler's paper that resulted in an article by Wright entitled ``On the Occurrence of the Enhanced lines of Nitrogen in the Spectra of novae" \citep{wrightnitflare}. It is noted there that the ``4640 stage" occurs first on entering the nebular stage and then can occur recurrently. 

In the data presented herein, the flaring episodes reoccur in stages corresponding to the cusps observed in the nova light curve during the transition 
from the auroral to nebular spectral stages. The proposition that N~{\sc iii} is responsible for this flaring episode is justified 
in \cite{introastro} by a decrease in [O~{\sc iii}] and the ``great width of N~{\sc iii} lines corresponding to a velocity of 
3200 km s$^{-1}$". In the observations, no decrease in [O~{\sc iii}] was witnessed, but instead an increase. Also, the 
large Doppler width is not necessary if the feature is assigned to the eight lines of the O~{\sc ii} V1 multiplet, see 
Table \ref{V1V2}. If these eight lines were fully resolved in the observations, further diagnostics could be conducted, 
as was done in \cite{storey}, except for higher density media. It must be noted that these results have only been shown 
for slower CO nova eruptions and the Bowen fluorescence mechanism may still be responsible for the ``4640" feature observed 
early after eruption in faster nova events as well as a feature present in this region during the late nebular stage of some novae.

Concentrating on the nova during its auroral spectral phase, multiple components of the nova system are observed simultaneously. A
s the dust shell clears, the central source is revealed, evident from the rise out of the dust-dip in the optical and appearance 
of the super-soft-source in X-rays. From the suggestions of multiple ejection episodes from the flat-top-jitter phase, internal 
shocks can be expected, leading to a fracturing of the shell into cold and dense clumps. The nova shell is already 
expected to have been a shaped bipolar structure, implying that the polar and equatorial outflow do not have common 
distances from the central ionising source, which continues burning the residual nuclear material remaining on the surface. 

While studying V356 Sgr (1936), \cite{mclaughOII} found that \cite{wrightnitflare} was probably mistaken in his derivation of 
N~{\sc iii} being the dominant component in the ``4640" blend. \cite{1975mcclintock} analysed the origin of the same emission lines 
where a dense 10$^{10}$ cm$^{-3}$ shell is ionised by the stellar super-soft X-ray component and collisionally. \cite{warner} 
states that during these spectral stages, the electron density of the visible gas is of the order 10$^{7}$ - 10$^{9}$ cm$^{-3}$. 
Under these density constraints, {\C} models reveal that the previously expected N~{\sc iii} lines do not appear due to Bowen 
fluorescence but instead are heavily dominated by the aforementioned O~{\sc ii} blend. With densities intermediate to those 
suggested by \cite{warner} and 
\cite{derdzinski}, the O~{\sc ii} V1 multiplet can account for all the emission seen peaking around the 
4640 - 4650 $\rm{\AA}$ region on day 145 post-discovery spectrum in Fig. \ref{fig:spec3} (bottom panel). 

Excess H$\gamma$ 
emission at 4340 $\rm{\AA}$ may come from the O~{\sc ii} V2 multiplet around 4340$\rm{\AA}$ as well as the [O~{\sc iii}] 4
363 $\rm{\AA}$ auroral line, see Fig. \ref{fig:spec5}. Supported in the presented observations as a decrease in the population of 
O~{\sc i} perceived along with a 
corresponding increase in the O~{\sc ii} and O~{\sc iii} species. The emission from N~{\sc iii} at 4640 $\rm{\AA}$ is 
associated with O~{\sc iii} emission in the UV as it is also pumped through Bowen fluorescence. The C~{\sc iii} 
line often associated with the 4650 $\rm{\AA}$ region begins to be important at lower densities than those considered here 
($<$ 10$^{7.7}$ cm$^{-3}$). The O~{\sc ii} V1 multiplet appears under high-density and low-temperature conditions, suggesting that the emission has its origins in a cool and dense shell. \cite{metzger} explored the conditions present in a nova outflow, concentrating on shocks, and \cite{derdzinski} compared the rate of change of density and temperature from regular expansion to expansion with the presence of shocks. The top two panels of Fig. 2 in \cite{derdzinski} compare well to the expected values from the photoionisation model grid displayed in Fig. \ref{fig:cloudy1} of this work. 

\renewcommand{\arraystretch}{0.8}
\begin{table}[H]
\small
\centering
\caption[List of wavelengths of V1 and V2 O~{\sc ii}]{List of wavelengths of V1 and V2 O~{\sc ii} multiplet 
wavelengths along with the lower and upper terms of their transitions respectively, from \cite{storey}. Possible blending 
lines are listed along with their A$_{ki}$ values from the NIST database.
The initial and final levels are given next to the line i.d.}
\label{V1V2}
\begin{tabular}{l|llll}
\toprule
i.d.   & \multicolumn{1}{c|} {Wavelength ( $\rm{\AA}$)} &  A$_{ki}$ (s$^{-1}$)     &            \\
\midrule
   
O~{\sc ii} V1&   \multicolumn{1}{c|} {2s$^{2}$2p$^{2}$($^{3}$P)3p $^{4}$D$\rm^{o}$} & 2s$^{2}$2p$^{2}$($^{3}$P)3s $^{4}$P$\rm^{e}$  \\
\hline
 & \multicolumn{1}{c|}{4638.86}  &     $3.61\times10^{7}  $                        \\
      & \multicolumn{1}{c|}{4641.81}    &    $5.85\times10^{7}$                       \\
       & \multicolumn{1}{c|}{4649.13}    &     $7.84\times10^{7}$                            \\
       & \multicolumn{1}{c|}{4650.84}    &   $6.70\times10^{7}$                        \\
       & \multicolumn{1}{c|}{4661.63}    &     $4.04\times10^{7}$                          \\
       & \multicolumn{1}{c|}{4673.73}    &     $1.24\times10^{7}$                   \\
       & \multicolumn{1}{c|}{4676.23}    &      $2.05\times10^{7}$                     \\
       & \multicolumn{1}{c|}{4696.35}    &     $3.15\times10^{6}$             \\
    \midrule
N~{\sc iii} &   \multicolumn{1}{c|}{2s$^{2}$3p $^{2}$D} & 2s$^{2}$3d$^{2} $P$\rm^{o}$  \\
      \hline
  & \multicolumn{1}{c|}{4634.14}  &   $6.36 \times10^{7}$   \\     
   & \multicolumn{1}{c|}{4640.64}  &    $7.60\times10^{7}$                       \\
       \midrule
C~{\sc iii} &   \multicolumn{1}{c|}{1s$^{2}$2s3s $^{3}$P$\rm^{o}$} & 1s$^{2}$2s3p $^{3}$S \\
      \hline
      & \multicolumn{1}{c|}{4647.42}  &    $7.26\times10^{7}$                       \\   
      & \multicolumn{1}{c|}{4650.25}  &    $7.25\times10^{7}$                       \\
      & \multicolumn{1}{c|}{4651.47}  &    $7.24\times10^{7}$                       \\
      \midrule
 He~{\sc ii} & \multicolumn{1}{c|} {3p $^{2}$S} &  4s $^{2}$P$\rm{^{o}}$\\
\hline
      & \multicolumn{1}{c|}{4685.90} & $1.95\times10^{7}$  \\
    \midrule
    \midrule
O~{\sc ii} V2      &    \multicolumn{1}{c|}{2s$^{2}$2p$^{2}$($^{3}$P)3p $^{4}$P$\rm^{o}$} & 2s$^{2}$2p$^{2}$($^{3}$P)3s $^{4}$P$\rm{^{e}}$ \\
      \hline 
 & \multicolumn{1}{c|}{4317.14} &            $3.68\times10^{7}$             \\
       & \multicolumn{1}{c|}{4319.63}    &  $2.48\times10^{7}$              \\
       & \multicolumn{1}{c|}{4325.76}    &     $1.42\times10^{7}$             \\
       & \multicolumn{1}{c|}{4336.86}    &     $1.53\times10^{7}$              \\
       & \multicolumn{1}{c|}{4345.56}    &     $7.95\times10^{7} $          \\
       & \multicolumn{1}{c|}{4349.43}    &      $6.75\times10^{7} $           \\
       & \multicolumn{1}{c|}{4366.89}    &     $3.92\times10^{7}$         \\
       \midrule
       H$\gamma$  &    \multicolumn{1}{c|}{2p $^{2}$S} & 5s $^{2}$P$\rm{^{o}}$ \\
       \hline
      & \multicolumn{1}{c|}{4341.70} & $1.29\times10^{6} $ \\ 
      \midrule
      {[}O~{\sc iii}{]}  &    \multicolumn{1}{c|}{2s$^{2}$2p$^{2}$ $^{1}$D} & 2s$^{2}$2p$^{2}$ $^{1}$S \\
      \hline
       & \multicolumn{1}{c|}{4363.21}    &     $1.71\times10^{0}$        
\end{tabular}
\end{table}

\section{Discussion}
\label{discussion}

The slow novae with observed polarisation and visible nova shells are DQ Her, HR Del, V705 Cas, T Pyx, FH Ser and LV Vul. 
These seven novae all share similarities with V5668 Sgr (2015) in terms of light curve shape and suspected white dwarf 
composition. It is even possible that DQ Her and V5668 Sgr coincidentally both have white dwarf spin periods of 71 s. It was 
found that densities of nova shells during the spectral stages studied in this work are poorly constrained and that the upper 
limit of 10$^{9}$ cm$^{-3}$ in \cite{warner} may be an underestimation. The observed nebular lines of  [O~{\sc iii}] are due to de-excitation after the auroral 4363 $\rm{\AA}$ transition. An analytical problem that arises in this work is that the plasma diagnostics from the literature are only applicable to lower density gas, such that further simulations are required to properly numerically reproduce and understand nova shells during these early stages of evolution post-eruption. 
Correct identification of observed lines are therefore of great importance and there is strong evidence that lines get systematically misidentified in the literature. Analysis is hampered by the blending of many lines in erupting nova systems, which is exacerbated by their large Doppler broadening. 

It is significant that an intrinsic change in absolute polarisation should be detected in the dataset, see \cite{evanspol} 
for a discussion on polarisation detection regarding novae. 
From a qualitative review of polarimetric 
studies of novae it is the slow novae that have the largest observed intrinsic change in polarimetric measurements over time.

Densities above 10$^{8}$ cm$^{-3}$ are rarely treated in novae, as higher 
densities, due to shock compression, have been recently called for to explain the observed gamma-ray emission from novae. 
This lack of treatment for shock compression early in an erupting nova's lifetime, see \cite{derdzinski} is because 
densities of the ejecta at this stage of the eruption were previously thought to be of the order of 10$^{6}$ cm$^{-3}$ 
and therefore within the normal nebular diagnostic limits. The nova shell of V5668 Sgr is expected to be photoionised 
by low-level nuclear burning on the white dwarf, peaking in the X-ray (BB peak 14 - 30 $\rm{\AA}$). The emission-line spectrum 
is dominated by permitted, auroral and nebular lines, e.g. O~{\sc i}, O~{\sc ii}, [O~{\sc iii}]. Referring to, 
\cite{williamsnot_MASON} where there is an in depth discussion on the optical depth of the [O~{\sc i}] 6300 + 6364 $\rm{\AA}$ lines, 
it is understood that high densities and strong radiation fields are responsible for their strength in novae. These lines are not 
well reproduced in the {\C} modelling presented in this work. It is thought that the [O~{\sc i}] lines originate in the same zones as the dust resides where densities are greatest. 
It is known that some lines are particularly sensitive to temperature (such as O~{\sc iii} transitions), whereas 
others are sensitive to density 
(O~{\sc ii} recombination), but this does not always hold true outside the normal nebular diagnostic limits.  

On the dust condensation timescale, a relation was derived by \cite{Williams:2013aa}, see their Fig. 2, where a comparison was made between a nova's  t$_{2}$ value and the onset of dust formation. Speed class relations are subject to scrutiny, see \cite{kasliwalbook}, especially in flat-top-jitter 
novae as they vary considerably in their early light curves, unlike their faster and smoother counterparts. It is therefore prescribed that the t$_{2}$ and t$_{3}$ values for this type of nova should be taken from their final drop in the early observed maxima, giving a value for V5668 Sgr of around 60 days for t$_{2}$. The relation from \cite{Williams:2013aa} gives an onset of dust formation at day 80, in accordance with the beginning of the deep dust-dip marked in Fig. \ref{fig:V5668Sgr_LC}. In \cite{evansv339del}, the relationship between the dust formation episode and the duration of the X-ray emission of V339 Del were studied where it was found that the end of the super-soft-source phase corresponded with the end of the strong dust-dip of the nova. This work found that the hard radiation field it is exposed to during the super-soft-source phase likely destroys dust. 

Gamma-ray emission from novae has been proposed to be intrinsically linked with the nova shell's geometry. 
In order to explain observed gamma emission from novae, shocks between a slow-dense-ejecta and a faster-chasing-wind 
appear necessary, e.g. \citet{finzell,cheung_gammav5668sgr}. 
The 55 day period over which gamma rays were detected for this nova in \citet{cheung_gammav5668sgr}, 
during the flat-top-jitters of Fig. \ref{fig:V5668Sgr_LC}, implies a lengthy cycle of shocks between the 
slow dense ejecta and fast chasing wind, possibly leading to strong shaping of the nova remnant. The current 
understanding of gamma-emission from classical novae suggests that a denser equatorial waist and lower density 
polar ejecta should exist for this nova, which is strongly supported by the polarimetric and spectroscopic 
observations presented in this work as well as in the NIR study conducted by \citet{banerjeeV5668sgr}.

Morpho-kinematic modelling of nova shells suggests that DQ Her-like novae are seen edge on and that the long eruption light curve 
is due to reprocessing of light in the dense outflow.

\section{Conclusions}
\label{conclusions}

The observations reveal variability of the absolute polarisation before and after nights that hint towards internal shocks 
in the nova outflow. Along with the available high-quality gamma, X-ray, UV and IR observations on this nova, the polarimetry 
allowed for the estimation of the nova shell position angle and provided information on the dust grains causing the scattering. 
The spectroscopy then allowed for derivation of the physical conditions on separate nights, including outflow velocity and 
structure, nebular density, temperature and ionisation conditions. Following on from this extensive analysis, morpho-kinematic 
and photoionisation models were formulated and combined to give a deeper insight into the nova system as a whole.
Finally we note that, for slow novae in particular, the regularly referred to `nitrogen flaring' is in fact more likely to be 
`oxygen flaring'.

\begin{acknowledgements}

A special thank you to Dr. Christophe Morisset and Prof. Iain Steele for invaluable discussions. The Liverpool Telescope is operated on the island of La Palma by Liverpool John Moores University (LJMU) in the Spanish Observatorio del Roque de los Muchachos of the Instituto de Astrof\'isica de Canarias with financial support from STFC. E. Harvey wishes to acknowledge the support of the Irish Research Council for providing funding for this project under their postgraduate 
research scheme. S. C. Williams acknowledges a visiting research fellowship at LJMU. The authors gratefully acknowledge with thanks the variable star observations from the AAVSO International Database contributed by observers worldwide and used in this research. 

\end{acknowledgements}
%
%


%

%


%

\begin{appendix}

\onecolumn

\begin{landscape}
\begin{table}[]
\centering
\setlength\tabcolsep{1.0pt}
\caption[H$\beta$ flux measurements and line ratios]{H$\beta$ flux measurements on the days after detection relevant to the ratios in 
table below. The flux measurements are in units of ergs/cm$^{-2}$/s/$\rm{\AA}$ and have not been corrected for extinction. 
In the long table below the ratio of observed line fluxes to that of $H\beta$ are presented over the same 10 nights covering 
days 114, 116, 120, 122, 123, 130, 141, 143, 145 and 153 post-discovery respectively. The relevant spectra can be seen in 
Figs. \ref{fig:spec1} and \ref{fig:spec2}. The H$\beta$ fluxes have been normalised to 100 from the values stated for the relevant dates. As H$\beta$ 
flux is dependent on the filling factor of the nova shell the quoted line ratios are sensitive to such. No reddening 
correction has been applied to the values in the following table. In the following table with line ratios the included model values are 
from a log of hydrogen density model of 9.0 and an effective temperature of $2.2\times10^{5}$ K.
}
\label{fluxl}
\begin{tabular}{|l|l|l|l|l|l|l|l|l|l|l|l|l|}
\hline
\multicolumn{1}{c|}{\textbf{${\lambda}$}} & \multicolumn{1}{c|}{\textbf{line i.d.}} & \multicolumn{1}{c|}{\textbf{114}}  & \multicolumn{1}{c|}{\textbf{116}} & \multicolumn{1}{c|}{\textbf{120}}  & \multicolumn{1}{c|}{\textbf{122}}  & \multicolumn{1}{c|}{\textbf{123}} & \multicolumn{1}{c|}{\textbf{130}}  & \multicolumn{1}{c|}{\textbf{141}}  & \multicolumn{1}{c|}{\textbf{143}}  & \multicolumn{1}{c|}{\textbf{145}}  & \multicolumn{1}{c|}{\textbf{153}}\\ \hline 
     \hline
\multicolumn{1}{c|}{\textbf{4861 $\rm{\AA}$}} & \multicolumn{1}{c|}{\textbf{H$\beta$}} & \multicolumn{1}{c|}{\textbf{2.2$\times10^{-13}$ }}  & \multicolumn{1}{c|}{\textbf{2.9$\times10^{-13}$}} & \multicolumn{1}{c|}{\textbf{2.6$\times10^{-13}$}}  & \multicolumn{1}{c|}{\textbf{2.9$\times10^{-13}$}}  & \multicolumn{1}{c|}{\textbf{5.3$\times10^{-13}$}} & \multicolumn{1}{c|}{\textbf{8.2$\times10^{-13}$}}  & \multicolumn{1}{c|}{\textbf{2.7$\times10^{-12}$}}  & \multicolumn{1}{c|}{\textbf{2.7$\times10^{-12}$}}  & \multicolumn{1}{c|}{\textbf{2.9$\times10^{-12}$}}  & \multicolumn{1}{c|}{\textbf{2.3$\times10^{-12}$}}\\ \hline 

\hline
\end{tabular}
\end{table}


\begin{longtable*}{|l|l|l|l|l|l|l|l|l|l|l|l|l|} 
\label{tablelongratios}
\\ \hline 
\multicolumn{1}{c|}{\textbf{${\lambda}$}} & \multicolumn{1}{c|}{\textbf{line i.d.}} & \multicolumn{1}{c|}{\textbf{114}}  & \multicolumn{1}{c|}{\textbf{116}} & \multicolumn{1}{c|}{\textbf{120}}  & \multicolumn{1}{c|}{\textbf{122}}  & \multicolumn{1}{c|}{\textbf{123}} & \multicolumn{1}{c|}{\textbf{130}}  & \multicolumn{1}{c|}{\textbf{141}}  & \multicolumn{1}{c|}{\textbf{143}}  & \multicolumn{1}{c|}{\textbf{145}}  & \multicolumn{1}{c|}{\textbf{153}} & \multicolumn{1}{c|}{\textbf{Mod}} \\ \hline 

\endfirsthead

{{\bfseries(continued)}} \\

\multicolumn{1}{c|}{\textbf{${\lambda}$}} & \multicolumn{1}{c|}{\textbf{line i.d.}} & \multicolumn{1}{c|}{\textbf{114}}  & \multicolumn{1}{c|}{\textbf{116}} & \multicolumn{1}{c|}{\textbf{120}}  & \multicolumn{1}{c|}{\textbf{122}}  & \multicolumn{1}{c|}{\textbf{123}} & \multicolumn{1}{c|}{\textbf{130}}  & \multicolumn{1}{c|}{\textbf{141}}  & \multicolumn{1}{c|}{\textbf{143}}  & \multicolumn{1}{c|}{\textbf{145}}  & \multicolumn{1}{c|}{\textbf{153}} & \multicolumn{1}{c|}{\textbf{Mod}} \\ \hline 

\endhead
\hline 
\multicolumn{2}{|r|}{{Continued on next page}} \\ \hline
\endfoot

\endlastfoot
4861       & H$\beta$               & 100  & 100 & 100 & 100 & 100 & 100 & 100 & 100 & 100 & 100 & 100 \\
3970       & H$\epsilon$            &      &     &     &     &     & 12  & 14  & 13  & 15  & 12  &   \\
4089       & OII                    &      &     &     &     &     & 2   & 7   & 5   & 10  & 3   &     \\
4102       & H$\delta$              & 28   & 23  & 27  & 27  & 28  & 26  & 30  & 30  & 36  & 27  & 35 \\
4176/4179  & FeII                   & 2    & 1   & 1   & 1   & 2   & 2   &     & 1   & 4   & 2   &     \\
4185/90    & OII                    &      &     &     &     &     & 2   &     &     & 4   & 2   &     \\
4200       & HeII                   &      &     &     &     &     &     & 3   &     & 5   & 2   &     \\
4233       & FeII                   & 6    & 5   & 4   & 6   & 5   & 2   & 3   & 2   & 3   & 2   &     \\
4267       & CII                    &      &     & 3   & 5   & 4   & 3   & 6   & 5   & 6   & 4   &     \\
4340       & OII/H$\gamma$          & 43   & 43  & 48  & 49  & 47  & 45  & 48  & 47  & 50  & 45  & 53  \\
4363       & [OIII]     &      &     &     &     &     &     & 11  & 9   & 12  & 8   &   17\\
4388       & HeI                    &      &     &     &     &     & 6   & 8   &     & 11  &     &     \\
4414/17  & OII/FeII               & 9    & 6   & 6   & 8   & 7   & 9   & 14  & 10  & 10  & 9   &     \\
4452       & [FeII]/OII & 5    & 5   & 5   & 7   & 5   & 8   & 12  & 6   & 8   & 6   &     \\
4471       & HeI                    & 2    & 1   & 2   & 4   & 2   & 4   & 8   & 6   & 9   & 5   &  4  \\
4515/20    & FeII                   & 1    & 1   & 1   & 1   & 2   & 8   & 8   & 5   & 14  & 6   &     \\
4523/42    & FeII                   & 1    & 1   & 1   & 1   & 1   & 6   & 6   &     & 10  & 5   &     \\
4556       & FeII                   &      &     &     & 2   & 1   & 2   & 5   &     &     & 4   &     \\
4584       & FeII                   & 3    & 2   & 1   & 1   & 2   & 3   & 3   &     &     & 3   &     \\
4607/21 & NII                    & 3    & 4   & 4   &     &     &     &     &     &     & 5   &     \\
4651       & OII                    & 5    & 6   & 6   & 11  & 11  & 39  & 58  & 46  & 79  & 40  & 55  \\
4686       & HeII                   & 1    & 1   & 1   & 2   & 2   & 12  &     &   &     &     &  2 \\
4713       & HeI                    & 2    & 2   & 1   & 2   & 2   &     &     &     &     &     &  1   \\
4755       & [FeIII]    &      & 1   & 1   & 1   & 1   & 2   & 3   & 1   &     &     &     \\
4803/10    & NII                    & 3    & 3   & 2   & 4   & 3   & 4   & 4   & 2   & 4   & 2   &     \\
4922       & HeI                    &      & 13  & 11  & 12  & 11  & 11  & 7   & 9   & 9   & 9   &    \\
4959       & [OIII]     & 4    & 4   & 6   & 8   & 7   & 9   & 11  & 10  & 12  & 10  & 10  \\
5001       & NII                    & Blnd &     &     &     &     &     &     &     &     &     &     \\
5007       & [OIII]     & 13   & 14  & 18  & 22  & 19  & 25  & 29  & 31  & 34  & 26  & 30  \\
5016       & HeI                    & 15   & 12  & 12  & 17  & 14  & 12  & 10  &     &     &   &    \\
5048       & HeI                    & 2    & 2   & 5   &     & 8   &     &     & 9   & 10  & 6   &    \\
5111       & FeIII                  & 2    & 1   &     & 2   & 2   &     &     &     &     &     &     \\
5142       & FeII                   &      & 2   & 1   &     &     &     & 3   &     & 3   &     &     \\
5167       & FeII                   & 7    & 7   & 6   & 8   & 7   & 6   & 5   & 3   & 4   & 4   &     \\
5176/80    & NII                    & 5    & 4   & 4   &     & 5   & 5   & 3   & 3   &     & 3   &     \\
5197       & FeII                   & 1    & 1   & 2   & 3   & 2   & 3   & 3   & 2   &     &     &     \\
5197/200   & [NI]/FeII  & 1    & 1   & 1   & 2   & 1   & 3   & 3   & 2   & 3   & 3   &     \\
5276       & FeII                   & 5    & 4   & 5   & 6   & 5   & 6   & 4   & 3   & 5   & 5   &     \\
5314       & FeII                   & 4    & 3   & 4   & 4   & 4   & 4   & 4   & 2   & 4   & 4   &     \\
5363       & FeII                   & 2    & 2   & 2   & 3   & 2   &     &     & 4   &     &     &     \\
5411       & FeII                   & 1    & 1   & 1   & 1   & 1   & 3   & 1   & 1   & 2   & 2   &     \\
5667/80    & NII                    &      &     &     &     &     & 12  & 13  & 9   & 15  & 12  &     \\
5876       & HeI                    &      &     &     &     &     &     & 7   & 13  & 15  & 10  &   \\
5940/42    & NII                    &      &     &     &     &     &     & 5   & 6   & 3   &     &     \\
6148       & FeII                   &      &     &     &     &     &     & 2   &     &     &     &     \\
6168       & NII                    &      &     &     &     &     &     & 2   &     &     &     &     \\
6300       & [OI]       & 125  & 108 & 82  & 100 & 78  & 45  & 31  & 28  & 27  & 24  &     \\
6364       & [OI]       & 45   & 38  & 29  & 36  & 28  & 16  & 12  & 9   & 10  & 9   &     \\
6482       & NII                    & 7    &     & 9   & 9   & 8   & 10  & 14  & 5   & 7   & 4   &     \\
6683       & HeII                   &      &     &     &     &     &     & 10  & 6   &     & 3   &     \\
6716       & [SII]      &      &     &     &     &     &     & 12  & 6   &     & 2   &     \\
6731       & [SII]      &      &     &     &     &     &     & 10  &     & 4   &     &     \\
7065       & HeI                    &      &     &     &     &     & 11  & 12  & 11  & 15  & 11  & 12   \\
7116/20    & CII                    &      & 7   & 9   & 13  & 14  & 11  & 11  & 6   & 11  & 9   &     \\
7236       & CII                    &      &     &     &     &     & 12  & 15  & 9   & 9   & 9   &     \\
7325       & OI                     & 55   & 58  & 55  & 59  & 52  & 50  & 48  & 50  & 54  & 48  & 45  \\
7442       & NI                     & 2    & 4   &     & 4   & 7   &     &     &     &     &     &     \\
7468       & NI                     & 3    & 4   &     & 6   &     & 8   &     &     &     &     &     \\
7725       & [SI]       &      &     &     &     &     &     &     &     & 6   &     &     \\
7772/5     & OI                     & 18   & 16  & 17  & 22  & 20  & 18  & 10  & 12  & 13  & 14  &     \\
7877/86    & MgII                   &      &     &     &     & 5   & 7   & 1   & 4   & 4   & 6   &     \\
8166       & NI                     &      &     & 3   & 6   & 8   & 12  & 1   &     &     &     &     \\
8211       & NI                     &      & 14  & 14  & 21  & 19  & 18  & 16  & 10  & 11  & 12  &     \\
8228       & OI                     & 17   & 6   & 5   & 10  & 11  & 13  & 12  & 7   & 9   & 9   &     \\
8335/46    & CI/HI                  &      &     &     &     &     & 11  & 11  & 5   & 8   & 7   &    \\
8446       & OI                     & 398  & 354 & 336 & 380 & 328 & 221 & 145 & 173 & 139 & 153 & 153 \\
8498       & HI                     &      & 7   &     &     &     &     & 8   & 4   & 5   & 8   &     \\
8545       & HI                     & 5    & 6   & 4   & 9   & 10  & 10  & 9   & 6   & 4   & 8   &    \\
8594/98    & HI                     & 13   & 10  & 13  & 15  & 13  & 12  & 12  & 7   & 8   & 9   &   \\
8629       & [NI]       & 20   & 15  & 11  & 21  & 19  & 13  & 10  & 7   &     & 9   &     \\
8665       & HI                     & 25   & 24  & 21  & 26  & 28  & 19  & 18  & 11  & 14  & 14  &   \\
8686       & NI                     &      &     & 9   & 14  & 14  & 11  & 10  & 12  &     &     &     \\
8750       & HI                     & 10   & 8   & 9   & 13  & 13  & 12  & 12  & 8   & 7   & 10  &   \\
8863       & HI                     & 9    & 9   & 13  & 13  & 13  & 12  & 11  & 8   & 9   & 11  &   \\
9015       & HI                     & 15   & 13  & 11  & 17  & 18  & 15  & 13  & 9   & 7   & 9   &   \\
9029/60    & [NI]       &      & 10  & 7   & 8   & 7   &     & 10  &     & 6   &     &     \\
9208       & NI                     &      &     &     & 16  & 16  & 13  & 9   & 4   & 6   &     &     \\
9229       & HI                     & 20   & 19  & 20  & 24  & 25  & 21  & 18  & 16  & 13  & 14  &   \\
\hline
\end{longtable*}

\end{landscape}

\twocolumn

\begin{figure*}[ht!]
\centering
\includegraphics[width=18cm]{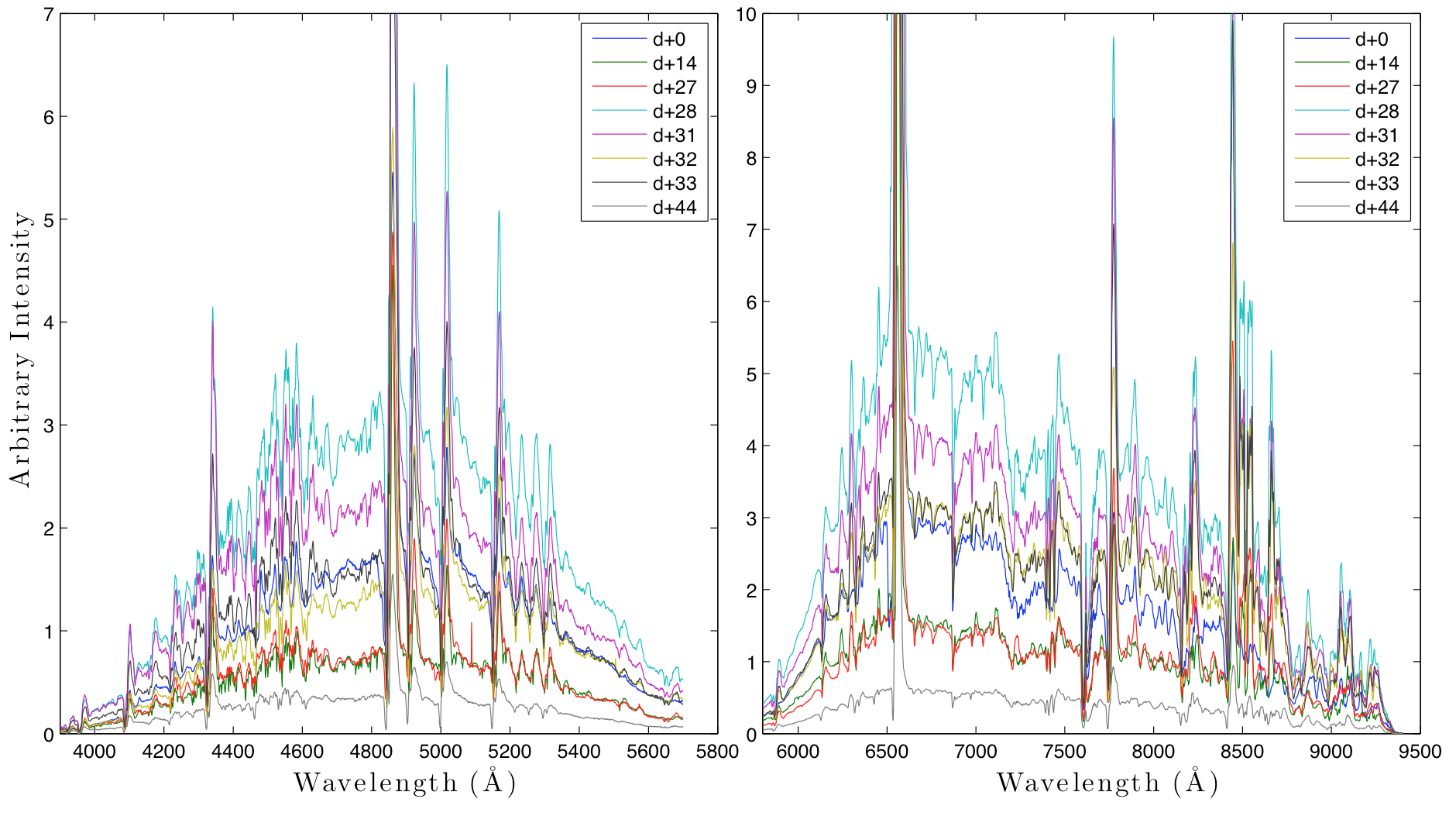}
\caption{Spectroscopy from the flat-top-jitter phase in the FRODOSpec blue and red arms. Dates post-discovery are marked on the upper right hand ride of the plot}
\label{fig:earlyspecqueue}
\end{figure*}

 \begin{figure*}[ht!]
\centering
\includegraphics[width=18cm]{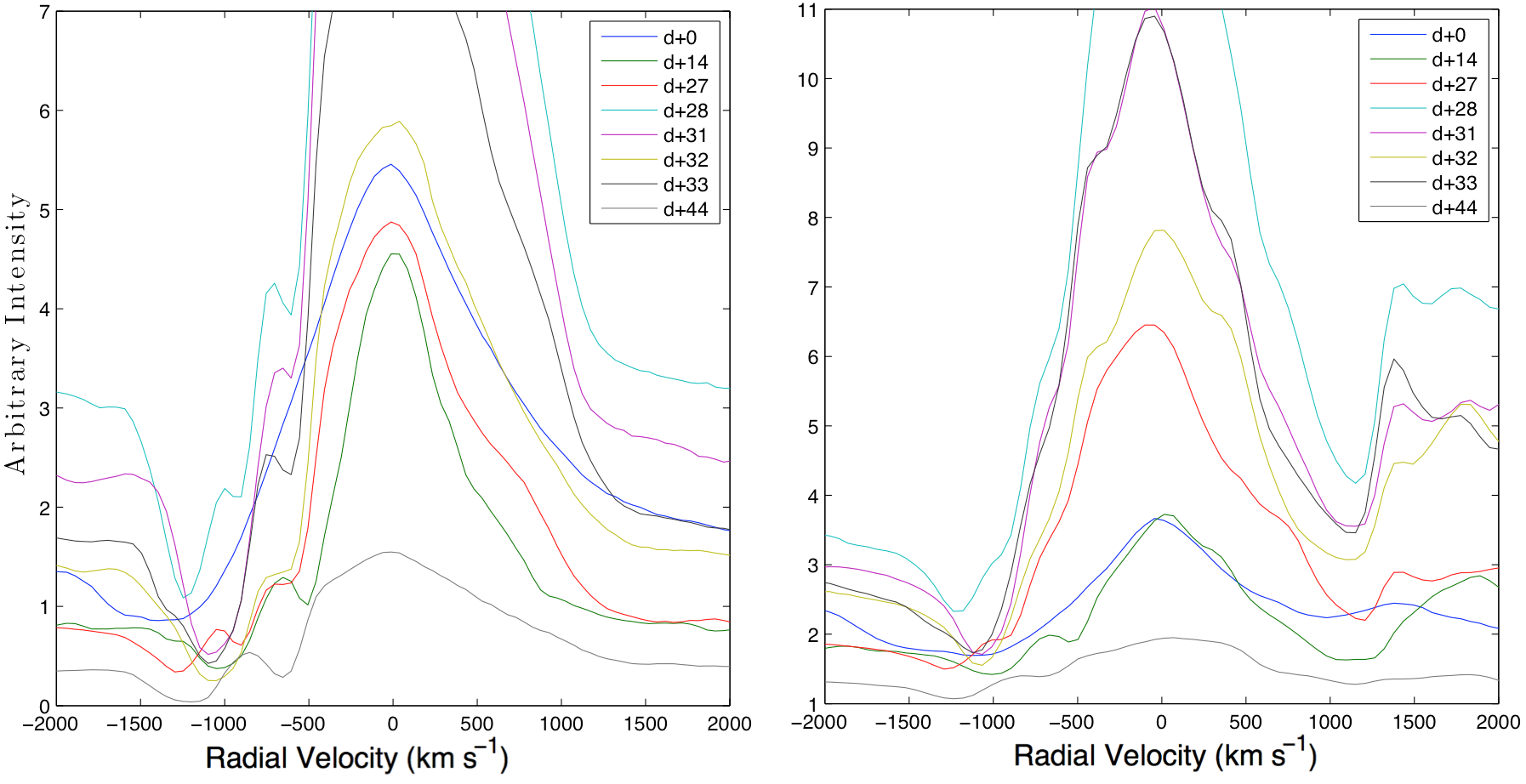}
\caption{Radial velocities of the flat-top-jitter epoch spectra from H$\beta$ on the left and O~{\sc i} on the right.}
\label{fig:earlyspec12wit}
\end{figure*}

\end{appendix}


\bibliographystyle{aa}
\bibliography{nova_nova_references}

\end{document}